# Tunneling and inversion symmetry in single-molecule magnets: the case of the $Mn_{12}$ wheel molecule

E. del Barco

*Department of Physics, University of Central Florida, Orlando, FL 32816, USA*

S. Hill

*National High Magnetic Field Laboratory and Department of Physics,*
*Florida State University, Tallahassee, FL 32310, USA*

C.C. Beedle and D.N. Hendrickson

*Department of Chemistry and Biochemistry, University of California at San Diego, La Jolla, CA 92093, USA*

I. S. Tupitsyn, P. C. E. Stamp

*Physics and Astronomy, and Pacific Institute of Theoretical Physics,*
*University of British Columbia, 6224 Agricultural Rd., Vancouver, BC, Canada V6T 1Z1*

We present a detailed study of the influence of various interactions on the spin quantum tunneling in a $Mn_{12}$ wheel molecule. The effects of single-ion and exchange (spin-orbit) anisotropy are first considered, followed by an analysis of the roles played by secondary influences, e.g. disorder, dipolar and hyperfine fields, and magnetoacoustic interactions. Special attention is paid to the role of the antisymmetric Dzyaloshinski-Moriya (DM) interaction. This is done within the framework of a 12-spin microscopic model, and also using simplified dimer and tetramer approximations in which the electronic spins are grouped in 2 or 4 blocks, respectively. If the molecule is inversion symmetric, the DM interaction between the dimer halves must be zero. In an inversion symmetric tetramer, two independent DM vectors are allowed, but no new tunneling transitions are generated by the DM interaction. Experiments on the $Mn_{12}$ wheel can only be explained if the molecular inversion symmetry is broken, and we explore this in detail using both models, focussing on the asymmetric disposition and rounding of Berry phase minima associated with quantum interference between states of opposite parity. A remarkable behavior exists for the 'Berry phase zeroes' as a function of the directions of the internal DM vectors and the external transverse field. A rather drastic breaking of the molecular inversion-symmetry is required to explain the experiments; in the tetramer model this requires a reorientation of the DM vectors on one half of the molecule by nearly 180°. This cannot be attributed to sample disorder. These results are of general interest for the quantum dynamics of tunneling spins, and lead to some interesting experimental predictions.

PACS numbers: 03.65.Yz, 75.45.+j, 75.50.Xx

## I: INTRODUCTION

The last decade or so has seen an explosion of interest in the tunneling dynamics of a large variety of magnetic molecules [1]. Throughout this period there has been an attempt on the experimental side to discern the properties of single molecule magnets (SMM), using the single molecule tunneling theory developed long ago [1–5] (with specific application to SMMs [6, 7] and later corrections for field-induced oscillation and Berry phase effects [8–11]). However this attempt has been confounded by the fact that unless applied fields in the system are very large, an independent tunneling model actually makes no sense: in reality each molecule is coupled strongly to the nuclear spin bath via hyperfine interactions [12, 13], to phonons via spin-phonon terms [13–18], and to other molecules via dipolar interactions [19, 20] (and occasionally also via superexchange terms). These couplings are typically far stronger than any low-field tunneling amplitude (unless strong fields are applied), so that even in the zero-temperature limit, single molecules must relax incoherently, i.e., the molecules typically do not tunnel independently at all [19, 20]. This was obvious, even in the very early experiments on magnetic molecule tunneling: the relaxation was both very slow and severely non-exponential [21–26], and the resonant hysteresis steps were extremely broad [21, 27, 28], with a width independent of the tunneling amplitude. It was also clear that dipolar and hyperfine interactions must strongly affect any experiment involving time-varying fields [29], such as the later Landau-Zener experiments [30] (and the theory since then [31–35] has made clear that this is a subtle problem: approximate solutions exist, valid in certain regimes, but there is no general solution so far). We note that these remarks apply to large-scale quantum dynamics in any magnetic system, whether one discusses tunneling domain walls [36–42], quantum spin glasses [43–48], or even room-temperature magnon BEC [49, 50]. In all of these cases, interactions with a spin bath [51] (and some-times with an oscillator bath [52, 53]) radically alter the single spin dynamics, and dipolar interactions then make the spin dynamics a collective process [54]. The effect of these interactions on coherent quantum dynamics in spin systems will be even more drastic, typically causing very strong decoherence [51, 55]. In the case of single-crystals containing SMMs, collective behavior typically governs the low-T dynamics of the molecular spins, as is the case, e.g., for tunneling-mediated long range dipolar ordering [56, 57], thermally-activated and tunneling-ignited magnetic avalanches [58–60], and random-field ferromagnetism [61], to mention a few examples.

In spite of these complications, it has been quite common for experiments on tunneling molecules to be interpreted in terms of a single molecule tunneling picture. However, severe inconsistencies can arise if this is done. Perhaps one of the most striking of these is the apparent violation of spin selection rules for single molecule tunneling. Peaks in the magnetic relaxation rate, interpreted as tunneling resonances, are seen not only at the applied fields expected from SMM tunneling theory, but also at other fields where they should have been forbidden by the molecular symmetry. Moreover, the extracted relaxation rates in the two classes of resonance are very similar (see, eg., refs. [26, 62]). One obvious possible explanation for this apparently systematic violation of 'spin selection rules' is sample disorder. This results first and foremost in a distribution of the microscopic Hamiltonian parameters describing the molecules due, e.g., to distinct molecular isomers [63–68] or strain fields [63]. These forms of disorder also typically give rise to small misalignments (tilts) of the molecules which, upon application of a longitudinal bias field, result in unavoidable random transverse fields [62, 64, 66–68].

The role of disorder can be hard to pin down in experiments, due to the many possible forms in a crystal containing large polynuclear magnetic molecules, e.g. ligand disorder, solvent disorder/loss, impurities, dislocations/strains, *etc.*. Nevertheless, it has been possible in a few cases to characterize and quantify the disorder, and to pin-down its very clear influence on the tunneling [64–68]. However, it is certainly not the case that disorder is the only factor responsible for the apparent violation of the spin selection rules. Intermolecular dipolar fields clearly also play a role, and these by their very nature involve multi-molecule collective effects.

Some recent experiments have been performed on systems in which disorder is very weak, allowing a new look at the above question. In fact experiments on the $Mn_3$ system have permitted the first clear observation of spin selection rules [69], i.e., an absence of tunneling at resonances forbidden according to SMM tunneling theory. This experiment also showed that the spin selection rules could be quite subtle: some (but not all) of the forbidden resonances can be '*switched on*' with a very weak transverse field, but only if one correctly orients the in-ternal Jahn-Teller axes of the $Mn^{3+}$ ions in the molecule. Consequently, one can see first hand how intermolecular dipolar fields can influence some of the forbidden single molecule tunneling transitions.

With all this in mind, the case of the $Mn_{12}$ wheel molecule then becomes unusually interesting. In this system, the unit cell is thought to possess a single molecule with an inversion center; this rigorously excludes tunneling between states having opposite parity under inversion. Nevertheless such inversion symmetry-breaking tunneling transitions were seen in this system by Ramsey et al. [70], with what appeared to be a well-defined tunneling rate. Ramsey et al. pointed out the contradiction with the inversion symmetry, and also noted that if one adopted a dimer model for the exchange coupling in this system, and then supposed that there was a small Dzyaloshinski-Moriya (DM) interaction between the 2 halves of the molecule, this might explain the tunneling. However, as they also emphasized, such a DM interaction, between the 2 dimer elements, is impossible if the molecule is inversion symmetric; only a breaking of this symmetry would permit this. They also discussed the possibility that this symmetry breaking could be caused by nuclear hyperfine fields, but argued that such interactions would be far too weak to explain the results. Thus the existence of the tunneling seemed rather mysterious.

In a later paper, Wernsdorfer et al. [71] saw essentially exactly the same as Ramsey et al. In spite of the symmetry argument just given, Wernsdorfer et al. nevertheless tried to interpret the results in terms of a well-defined tunneling rate for a single dimer, coming from a single DM interaction between the 2 dimer halves. Since then there has been an exchange of comments which re-iterate their opposing points of view [72–75], and the same issue has arisen in other molecules (see, for example, [76]).

This debate raises a number of specific questions about the $Mn_{12}$ system, as well as more general questions about the role of symmetry-breaking interactions in magnetic molecules. It also, yet again, raises the question of the circumstances under which one is justified in interpreting the data using naive models of independent single molecule tunneling. In this paper we will address these questions, focussing specifically on the $Mn_{12}$ wheel system, and on the way in which experiment and theory are related for this system. We pay particular attention to the role of the DM interaction, insofar as it exists; and we discuss the role of the other main interactions that intervene, notably dipolar and hyperfine interactions, and the possible effect of disorder.

One problem at the moment with the $Mn_{12}$ wheel system is that most of the internal couplings are unknown. Thus we have to extract many of the parameters from the experiments described here, and this means that theoretical models of these experiments, suggested by the experiments themselves, play an uncomfortably large role in their interpretation. To alleviate this situation some-what, we have discussed the results not only in terms of the dimer model that was introduced specifically for these systems [70, 77], but also in terms of a tetramer model. This latter model describes the molecule as composed of 4 parts whose mutual exchange couplings are quite weak. We emphasize immediately that at present there is not much evidence that this is a good model for the Mn$_{12}$ wheel system (although this may come in the future). Indeed, none of these models should be taken as the definitive theory for explaining the Mn$_{12}$ wheel. The dimer model works in terms of accounting for some of the QTM features observed at low temperatures, where only the lowest lying states intervene. And, as we show in this article, the tetramer model turns out to have many interesting properties in its own right, and it also illustrates very nicely the relation between local DM interactions and QTM in a molecule with global inversion symmetry. It is clear to the authors that all the couplings between the twelve manganese ions of the wheel would have to be taken into consideration to rigorously explain the intricate behavior of this molecule, as observed with other characterization techniques, such as EPR [78].

The paper is organized as follows:

In section II we give a discussion of the interactions in the Mn$_{12}$ wheel molecule, including the primary interactions (exchange, anisotropy, and DM interactions), and certain secondary interactions (dipolar, hyperfine, and magnetoacoustic). In section III we introduce the dimer and tetramer models, again discussing the different interactions in these descriptions. In both sections II and III, attention is also paid to the internal symmetries of the molecule and to the effects of disorder on the coupling strengths. In section IV we compare these models with experiment. We first review salient features of the experiments and their interpretation, and then discuss quantitatively how both the dimer and tetramer models behave under a variety of conditions. This involves some exploration of the parameter space of these models, with interesting results. We find that no explanation of the experiments is possible without a rather strong inversion symmetry-breaking, inexplicable by simple disorder. Finally, in section V, we conclude the discussion and suggest future directions for research.

## II: INTERACTIONS IN THE MN$_{12}$ WHEEL MOLECULE

In what follows we begin by discussing what is known about single Mn$_{12}$ wheel molecules, stressing the importance of the relationship between the symmetric and antisymmetric parts of the exchange interaction, and of interactions due to dipolar and hyperfine fields. We then discuss the 'dimer approximation' of the full single molecule Hamiltonian, and introduce a new 'tetramer' model, which goes one step beyond the dimer approximation. We go on to discuss the correct model to describe a set of interacting molecules.

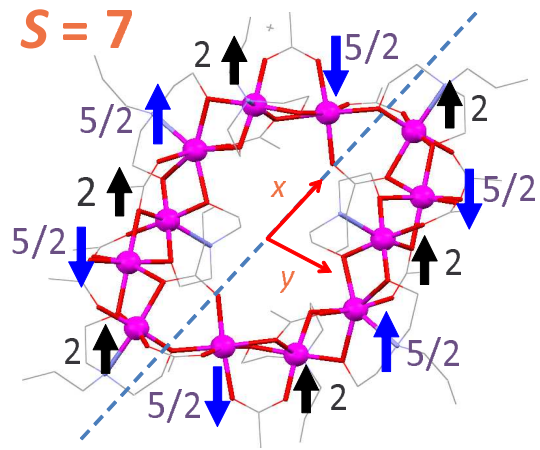

FIG. 1: (Color on-line) Structure of the [Mn$_{12}$Adea$_8$(CH$_3$COO)$_{14}$] molecule used for the experiments in ref. [70]. The dashed line indicates the magnetic separation of the wheel into two equal halves of spin $S = 7/2$, which, as discussed in the text, are ferromagnetically coupled to give a total spin $S = 7$ at low temperature.

### II.A: Dominant Interactions

Let us consider specifically the Mn$_{12}$ SMM studied experimentally in Ref. [70]. It consists of an alternation of six Mn$^{2+}$ ($s = 5/2$) and six Mn$^{3+}$ ($s = 2$) ions forming a single-stranded wheel [77, 79]. These wheels have the general chemical formula [Mn$_{12}$Rdea$_8$(CH$_3$COO)$_{14}$].$n$(CH$_3$CN), in which the Rdea$^{2-}$ are dianions of $N$-R diethanolamine, with R = A (allyl), b (butyl), e (ethyl), or m (methyl). The number $n$ of acetonitrile solvent molecules varies; for the experiments reported in ref. [70], $n = 7$ and the dianion Adea was used.

The structure of the [Mn$_{12}$Adea$_8$(CH$_3$COO)$_{14}$] molecule is shown in Fig. 1 (hereon referred to as the Mn$_{12}$ wheel). X-ray data [80] taken at 100 K show that the molecules crystallize in the $P\bar{1}$ space group, with a single molecule per unit cell. The unit cell dimensions are (using standard notation):

$$a = 16.0122(11)\text{Å} \quad \alpha = 63.1040(10)°$$
$$b = 16.3029(11)\text{Å} \quad \beta = 66.1200(10)°$$
$$c = 16.5860(11)\text{Å} \quad \gamma = 61.0910(10)° \qquad (1)$$

with a unit cell volume of $3282.7(4)$Å$^3$. Each molecule is symmetric under inversion. It is conceivable that this inversion symmetry is broken below 100 K. We will discuss this possibility later in the paper, but emphasize that so far there is no evidence for this symmetry-breaking from either magnetic or specific heat measurements con-ducted below 100 K (though we note that measurements were performed on powders and the specific heat data were of a low quality [80]).

### II.A.1: Microscopic Hamiltonian

A very simple model Hamiltonian for the $Mn_{12}$ wheel molecule combines the $Mn \cdots Mn$ superexchange couplings, the individual single-ion anisotropies, and the local Zeeman interactions, in a Hamiltonian of form:

$$H_0^{SM} = \sum_{j=1}^{12} [k_j^{\alpha\beta} s_j^\alpha s_j^\beta + O(s_j^4)] - \mu_B s_j^\alpha g_j^{\alpha\beta} B_0^\beta$$
$$+ \sum_{i<j} [J_{ij}^{\alpha\beta} + \epsilon_{\gamma\alpha\beta} D_{ij}^\gamma] s_i^\alpha s_j^\beta. \quad (2)$$

where $\epsilon_{\gamma\alpha\beta}$ is the unit antisymmetric tensor.

Here we sum over the 12 Mn sites, located at points $\mathbf{r}_j$ in space. As we discuss below, this Hamiltonian is probably only meaningful in the temperature range below a UV cutoff $\Omega_0 \sim 5$ K. The $k_j^{\alpha\beta}$ are the lowest-order single-ion anisotropy coefficients, the $J_{ij}^{\alpha\beta}$ the superexchange terms, and the $D_{ij}^\gamma$ are DM vectors (with $|\mathbf{D}_{ij}|$ having units of energy). The Zeeman interaction of each ion with the external field $\mathbf{B}_0$ is mediated by a $g$-tensor $g_j^{\alpha\beta}$. Note that the axes here are in spin space, and have no necessary connection with real space crystal axes - indeed, the 3 unit vectors $\{\mathbf{e}_j^\alpha\}$ (with $\alpha = x, y, z$), which define the axes in spin space at site $j$, are in general different for each one of the 12 Mn sites (although related in pairs by inversion).

*Some General Remarks*: All of the terms in (2) are produced by truncating out higher-energy atomic interactions. These include: hopping, on-site and near-neighbor Coulomb interactions, spin-orbit coupling, Hund's coupling, and charge transfer terms between $d-$ and $p-$ orbitals in a generalized Anderson model [81]. Since for transition metal-based systems the spin-orbit coupling term $\lambda_j \mathbf{l}_j \cdot \mathbf{s}_j$ is typically much weaker than the crystal field splittings $\Delta_j^{CF}$, we can meaningfully classify the terms in (2) by their order in $\lambda$; for Mn ions $\lambda_j \sim 300 - 500$ K, so that $\lambda/\Delta_j^{CF} \sim 0.02 - 0.06$. For a collection of 12 spins, having spin quantum numbers 2 and 5/2, a microscopic derivation of all terms in (2) from an Anderson model lies well beyond existing computational methods. Most of what we presently know about this molecule (and indeed any of the other large spin magnetic molecules) has come from a combination of experiments and phenomenological theory, along with some evidence from numerical work. We come to this below, but we first make a few more general theoretical remarks.

(a) It is known that the superexchange and DM terms in (2) are not independent (although this was not real-

ized in the early DM literature [82, 83]). The relationship between the two was completely characterized for Hamiltonians in which the superexchange is mediated by the hopping of single holes or electrons, via the $p$ orbitals of the bonding oxygens [84, 85]. One can then write the inter-spin interaction in the form of an *isotropic* exchange between "pseudospins" $\tilde{s}_l^\alpha$, viz.:

$$H_{ij}^0 = [J_{ij}^{\alpha\beta} + \epsilon_{\gamma\alpha\beta} D_{ij}^\gamma] s_i^\alpha s_j^\beta$$
$$\rightarrow \bar{J}_{ij} \tilde{s}_i^\alpha \tilde{s}_j^\alpha, \quad (3)$$

where $\bar{J}_{ij}$ is a renormalized isotropic superexchange coupling, and the pseudospins are rotated away from the original spins. This implies a hidden symmetry in the problem, so that the anisotropic part of the original $J_{ij}^{\alpha\beta}$ cannot be independent of $D_{ij}^\alpha$. Indeed, if we write the original interaction $J_{ij}^{\alpha\beta}$ in the form

$$J_{ij}^{\alpha\beta} = J_{ij}^0 \delta^{\alpha\beta} + \delta J_{ij}^{\alpha\beta}(1 - \delta^{\alpha\beta}), \quad (4)$$

then we must have

$$J_{ij}^0 = \left[ J_{ij}^{\gamma\gamma} - \frac{|D_{ij}^\gamma D_{ij}^\gamma|}{2J_{ij}^{\gamma\gamma}} \right]$$
$$\delta J_{ij}^{\alpha\beta} = \frac{D_{ij}^\alpha D_{ij}^\beta}{2J_{ij}^{\gamma\gamma}}. \quad (5)$$

We note that since $|\mathbf{D}_{ij}/J_{ij}^0|$ is formally $\sim O(\lambda)$, then $\delta J_{ij}/J_{ij}^0 \sim O(\lambda^2)$. The above expressions are more transparent if we define the spin basis so that for a given pair of sites $\{i, j\}$, the spin $\hat{z}$-axis is along the DM vector $\mathbf{D}_{ij}$. One can then write the total superexchange interaction in the form [86]

$$H_{ij}^0 = J_{ij}^0 [s_i^z s_j^z + \frac{1}{2} \cos\theta_{ij} (s_i^+ s_j^- + s_i^- s_j^+)$$
$$+ \frac{i}{2} \sin\theta_{ij} (s_i^+ s_j^- - s_i^- s_j^+)], \quad (6)$$

where $\tan\theta_{ij} = |\mathbf{D}_{ij}/J_{ij}^0|$ is the angle through which the hopping electron/hole spin is forced to rotate about the vector $\mathbf{D}_{ij}$ in passing from site $i$ to site $j$, this rotation being caused by the difference in local spin-orbit coupling between the 2 sites.

As is well known, all this means that: (i) in principle all bonds may have a finite $\mathbf{D}_{ij}$ (and corresponding anisotropy in $J_{ij}^{\alpha\beta}$, fixed by the magnitude and direction of $\mathbf{D}_{ij}$), but differing from one bond to another; however (ii) any internal symmetries in the superexchange links will be obeyed by the $\mathbf{D}_{ij}$ and the $J_{ij}^{\alpha\beta}$. The general implications of these points for transition metal systems in which the superexchange is mediated by single electrons or holes, are detailed by Yildirim et al. [85].

(b) There are other microscopic contributions to $H_0^{SM}$ which undermine the above arguments. First of all, none of these symmetry arguments apply rigorously for more general kinds of superexchange, involving higher-multiplicity spins (although in many cases they are a good guide). For the higher spin Mn ions we deal with here, the number of relevant configurations, and the number of symmetric and antisymmetric exchanges between them, becomes extremely large. No analysis at this level has ever been done, even for a single superexchange link, nor would it be practically useful. Second, as noted by Yildirim et al. [85], even for spin-1/2, the above arguments ignore off-site Coulomb and "Coulomb exchange" interactions. Third, as we discuss below, there exists a coupling between exchange interactions and phonons. Fourth, a finite field in the system can induce extra field-dependent magnetostrictive contributions to $H_0^{SM}$. Fifth and finally, any disorder in the system will induce a number of other terms, which we discuss below. All of these extra contributions give terms in $H_0^{SM}$ which violate the internal symmetry embodied in (3)-(6).

(c) Previous discussions of the single-ion anisotropy terms in (2) have assumed a simple quadratic biaxial form, writing [70, 87]

$$k_j^{\alpha\beta} s_j^{\alpha} s_j^{\beta} = -d_j s_{jz}^2 + e_j(s_{jx}^2 - s_{jy}^2),\tag{7}$$

where typically $d_j$ is the largest term, with higher-order terms in $\mathbf{s}_j$ omitted. Since what follows does not depend on the precise form of the single-ion anisotropy, we will use this form as well, just to be specific.

### II.A.2: Parameters for the Mn$_{12}$ wheel

There are two sources of information about the parameters in (2) for the case of the Mn$_{12}$ wheel. The first involves numerical evaluation of the exchange parameters $J_{ij}^0$ using local density functional (LDF) calculations. This has been done using a variety of methods [77, 87]. The results are simplified by the inversion symmetry: we only have to consider six nearest-neighbor bonds, and six next nearest-neighbor bonds (compare Fig. 2). If one assumes isotropic exchange, we can write the exchange part of the Hamiltonian $H_0^{SM}$ in the form

$$
\begin{aligned}
H_0^{exch} = &\sum_{j=1}^{6} J_j^0[\mathbf{s}_j \cdot \mathbf{s}_{j+1} + \mathbf{s}_{j+6} \cdot \mathbf{s}_{j+7}] \\
&+ \sum_{j=1}^{6} J_j^1[\mathbf{s}_j \cdot \mathbf{s}_{j+2} + \mathbf{s}_{j+6} \cdot \mathbf{s}_{j+8}],
\end{aligned}\tag{8}
$$

where a positive value denotes an antiferromagnetic coupling; note that this is opposite to the convention used in Ref. [87]. The distance $r_{ij}$ between the nearest-neighbor Mn ions varies from $r_{23} = 3.149$Å (for the link between sites 2 and 3, mediated by $J_2^0$), to $r_{67} = 3.473$Å, mediated by $J_6^0$. We can summarize the numerical results of refs. [77, 87] as follows:

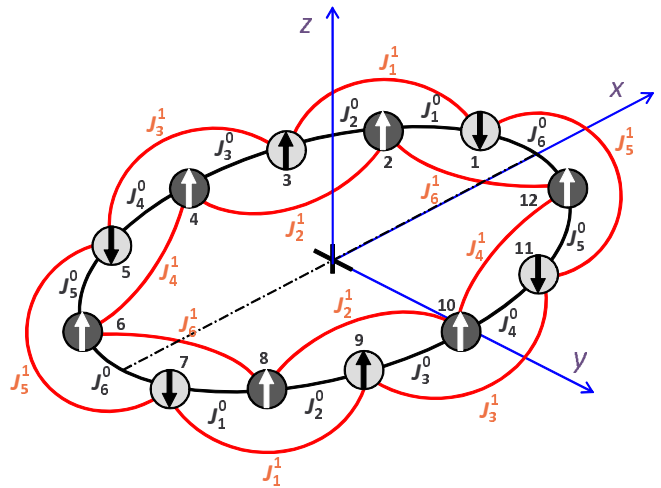

FIG. 2: (Color on-line) Labeling of the superexchange interactions in the Mn$_{12}$ wheel, including the nearest-neighbor interactions $J_j^0$, and 2nd nearest-neighbor interactions $J_j^1$ (cf. eqn. (8)); the values for these are discussed in the text. We assume inversion symmetry about the central point; the dashed line shows where the molecule is cut in the dimer *ansatz*.

(i) Most calculations find the nearest-neighbor parameter $J_6^0$ to be weak (results range from +0.9 to +1.8 cm$^{-1}$ using the Perdew-Burke-Erzenhof (PBE) functional, and from +7.0 to +7.3 cm$^{-1}$ using the B3LYP functional, depending on the spin configurations used; and a dinuclear approximation gives ~0 within computational error). The PBE and dinuclear (but not the B3LYP) results are consistent with the dimerized spin structure found in experiments, which requires that $|J_6^0|$ be considerably weaker than the other nearest-neighbor parameters. It is also argued [87] that $J_6^0$ must be antiferromagnetic, otherwise the system will not have an $S = 7$ ground state.

(ii) The $J_3^0$ and $J_4^0$ interactions are ferromagnetic, and almost identical, with values ranging from: −6.4 to −8.1 cm$^{-1}$ (PBE); −4.5 to −5.1 cm$^{-1}$ (B3LYP); and a dinuclear approximation gives −7 and −8 cm$^{-1}$, respectively.

(iii) The other nearest-neighbor parameters are all antiferromagnetic. However, the calculated values vary radically depending on the method used. For $J_1^0$, values range from +11.4 to +12.1 cm$^{-1}$ (PBE), from +4.1 to +4.4 cm$^{-1}$ (B3LYP), and a dinuclear approximation gives +2.8 cm$^{-1}$. For $J_2^0$, values range from +16.3 to +18.3 cm$^{-1}$ (PBE), from +7.2 to +8.0 cm$^{-1}$ (B3LYP), and a dinuclear approximation gives +9.2 cm$^{-1}$. For $J_5^0$, one finds +8.4 cm$^{-1}$ (PBE), a range from +2.8 to +5.4 cm$^{-1}$ (B3LYP), and a dinuclear approximation gives +5.0 cm$^{-1}$.

(iv) Not all of the next nearest-neighbor exchange parameters are small; $J_1^1$ is found to be +1.94 cm$^{-1}$ (PBE) or +0.7 cm$^{-1}$ (B3LYP); and $J_3^1, J_4^1, J_5^1$ are not much smaller. Particularly important is $J_5^1$, found to be +0.77 cm$^{-1}$ (PBE) or +0.27 cm$^{-1}$ (B3LYP); this cou-pling extends between the 2 dimer halves, and thus seriously competes with the small nearest-neighbor coupling $J_6^0$.

(v) The difference in energy between the $S = 7$ state and the other spin manifolds is not large (the gap between the $S = 7$ ground state and the lowest $S = 8$ state is estimated theoretically [87] to be $\sim 10$ cm$^{-1}$, and experiments find a gap to an $S = 6$ state of only 3.6 cm$^{-1}$ [70, 78]).

(vi) Since spin-orbit coupling was not incorporated, no results for either the single-ion anisotropy or the DM vectors were calculated.

We remark that it is not surprising that the results of these calculations depend quite strongly on the methods used. It is well known that for transition metal compounds, LDF theory does not give a terribly accurate guide to the effect of strong Coulomb correlations [81, 88], and so calculations of the $J_{ij}$ are fraught with uncertainty. For those parameters depending on spin-orbit and crystal field coupling, the range $\lambda/\Delta^{CF} \sim 0.02 - 0.06$ typical of the electrons in Mn ions implies that the DM interaction strength in each superexchange link will be $d_e \sim 0.4 - 1$ K per electron spin for the strongest of these links, and more like $d_e \sim 0.2 - 0.5$ K per electron spin for the weaker ones. Since $|\mathbf{D}_{ij}| = d_e/4 s_i s_j$ and all of the exchange links in the ring have $s_i s_j = 3$, we find that $|\mathbf{D}_{ij}| \sim 0.1 - 0.3$ K for the strongest of these links, and $\sim 0.07 - 0.15$ K for the weaker ones. Furthermore, this means that the exchange anisotropy will be extremely weak, at least 3 orders of magnitude smaller than $J^{ij}$, i.e., $\delta J \sim 1 - 10$ mK. The inversion symmetry of the wheel (assuming it is not broken) implies that we may write the DM terms as

$$H_0^{DM} = \sum_{j=1}^{6} \mathbf{D}_{j,j+1}^0 \cdot [(\mathbf{s}_j \times \mathbf{s}_{j+1}) + (\mathbf{s}_{j+6} \times \mathbf{s}_{j+7})] \quad (9)$$

so that DM vectors on opposite sides of the molecule are equal; however, each link has a rather low symmetry, so one cannot determine the directions of the $\mathbf{D}_{ij}^0$ by symmetry considerations alone.

The other source of information on all of these couplings is experimental. The results of Ramsey et al. [70] actually gave a somewhat different picture of the system from the numerical calculations; we discuss this experimental picture in sections III and IV below.

## II.B: Secondary interactions

A Hamiltonian like $H_0^{SM}$, containing only exchange and anisotropy terms, neglects some important interactions. Amongst these are the dipolar, hyperfine, and spin-phonon couplings, as well as 'extrinsic' effects from the combination of disorder and applied fields.

### II.B.1: Dipolar, hyperfine, and magnetoacoustic terms

A more realistic Hamiltonian for the SMM has the form

$$\mathcal{H}^{SM} = H_0^{SM} + H_1^{SM}, \quad (10)$$

with the term $H_1^{SM}$, viewed as a perturbation on the $H_0^{SM}$ given in (2) above, taking the form:

$$H_1^{SM} = \sum_{i<j} V_{ij}^{\alpha\beta} s_i^\alpha s_j^\beta + \sum_{jk} A_{jk}^{\alpha\beta} s_j^\alpha I_k^\beta + \sum_{j\mathbf{q}} \upsilon(\mathbf{s}_j, \phi_{\mathbf{q}}). \quad (11)$$

The first two terms in (11) come from integrating out the coupling of the electron and nuclear spin moments to the photon field. The dipolar interaction $V_{ij}$ between individual Mn spins has the usual form

$$V_{ij}^{\alpha\beta} s_i^\alpha s_j^\beta = \frac{\mu_0}{4\pi} \left[ \frac{\mathbf{m}_i \cdot \mathbf{m}_j}{r_{ij}^3} - 3 \frac{(\mathbf{m}_i \cdot \mathbf{r}_{ij})(\mathbf{m}_j \cdot \mathbf{r}_{ij})}{r_{ij}^5} \right], \quad (12)$$

in which the individual spin moments are

$$m_i^\alpha = \mu_B g_j^{\alpha\beta} s_j^\beta. \quad (13)$$

If we write $V_{ij}^{\alpha\beta}$ as

$$V_{ij}^{\alpha\beta} = V_{ij}^0 \mathbb{D}_{ij}^{\alpha\beta}, \quad (14)$$

where

$$V_{ij}^0 = \frac{\mu_0}{\pi} \frac{\mu_B^2}{r_{ij}^3}$$

$$\mathbb{D}_{ij}^{\alpha\beta} = \frac{g_i^{\gamma\alpha} g_j^{\delta\beta}}{4} \left[ \delta_{\gamma\delta} - \frac{3}{r_{ij}^2} r_{ij}^\gamma r_{ij}^\delta \right], \quad (15)$$

then $V_{ij}^0$ defines the energy scale of the dipolar interaction between the $i$-th and $j$-th spins (note that because of the angular factor in $\mathbb{D}^{\alpha\beta}$, the actual range of energies spanned by the dipolar interaction is actually $3V_{ij}^0 s_i s_j$, where $s_i = |\mathbf{s}_i|$, depending on the mutual orientation of the two spins).

The hyperfine couplings $A_{jk}^{\alpha\beta}$ exist between the 12 Mn electronic spin moments and (i) the 12 Mn nuclear spins, and (ii) the many other nuclear spins in the molecule (in the present case, there are 7 N and 63 H nuclei, plus the N and H nuclei in the 8 R dianion groups, and one can in principle also substitute finite spin isotopes of C and O).

Finally, there is a spin-phonon coupling $v(\mathbf{s}_j, \phi_{\mathbf{q}})$ at each spin site. It takes the form

$$v(\mathbf{s}_j, \phi_{\mathbf{q}}) = -\sum_{\alpha\beta\gamma\delta} \mathbb{A}_j^{\alpha\beta\gamma\delta} s_j^\gamma s_j^\delta \, u_j^{\alpha\beta} + O(u_j^2), \quad (16)$$

where the strain tensor $u_j^{\alpha\beta}$ is given in terms of the phonon displacement field $x_j^\alpha$ at position $\mathbf{r}_j$ by

$$u_j^{\alpha\beta} = \frac{1}{2} \left( \frac{\partial x_j^\alpha}{\partial r_j^\beta} + \frac{\partial x_j^\beta}{\partial r_j^\alpha} \right). \quad (17)$$

The coupling $\mathbb{A}_j^{\alpha\beta\gamma\delta}$ is the lowest order spin-phonon coupling, and formally $\mathbb{A}_j \sim O(\lambda)$, i.e., of the same order as the DM interaction (our separation of interactions into primary and secondary is thus a little arbitrary). Microscopic calculations of $\mathbb{A}$ (i.e., of the microscopic spin-phonon couplings) show that it is a combination of the spin anisotropy coefficients. In the present case we can estimate

$$|\mathbb{A}_j^{\alpha\beta\gamma\delta}| \sim |d_j|, \quad (18)$$

using (7), and assuming that the easy axis anisotropy dominates.

When we integrate out the phonons, down to the energy scale of the UV cutoff $\Omega_0$, we generate a magnetoelastic coupling between the Mn ions, of form

$$H_{int}^{ME} = \delta J_{ij}^{\alpha\beta\gamma\delta} s_i^\alpha s_i^\beta s_j^\gamma s_j^\delta, \quad (19)$$

as well as contributions, which we will ignore, to the fourth-order single-ion anisotropy terms $\delta k_i^{\alpha\beta\gamma\delta} s_i^\alpha s_i^\beta s_i^\gamma s_i^\delta$. Both $\delta J_{ij}^{\alpha\beta\gamma\delta}$ and $\delta k_i^{\alpha\beta\gamma\delta}$ are $\sim O(\lambda^2)$. The coupling constant $\delta J_{ij}^{\alpha\beta\gamma\delta}$ has a strength

$$|\delta J_{ij}^{\alpha\beta\gamma\delta}| \sim \frac{|\mathbb{A}_i \mathbb{A}_j|}{\rho c_0^2} \frac{1}{r_{ij}^3}, \quad (20)$$

and if we use (18), this gives $|\delta J_{ij}^{\alpha\beta\gamma\delta}| \sim |d_i d_j|/\rho c_0^2 r_{ij}^3$.

The numerical size of the terms in $H_1^{SM}$ can now be estimated for the Mn$_{12}$ wheel molecule:

(i) *The dipolar couplings*: If we assume a typical nearest-neighbor Mn distance of $\sim 3.2$ Å within the molecule, then the magnitude of the dipolar interaction is $V_{ij}^0 \sim 0.076$ K, assuming $g = 2$. These nearest neighbors have spin values $s = 2$ and $5/2$ respectively, so that $V_{ij}^{\alpha\beta} s_i^\alpha s_j^\beta$ spans an energy range $\sim 1.15$ K, depending on the mutual orientation of the two spins (we note that the next nearest-neighbor contributions are nearly 10 times smaller). It should be noted that the form of the intramolecular dipolar (spin-spin) coupling is such that it is virtually impossible to distinguish from the effects of the single-ion anisotropy terms $k_j^{\alpha\beta} s_j^\alpha s_j^\beta$ in any experiment, i.e. both interactions project in the same way onto the total molecular zero-field splitting

(anisotropy) tensor [89]. For this reason, it is common to take the view that intramolecular dipolar interactions are absorbed into eqn. (2). In contrast, eqn. (2) does not capture the effects of intermolecular dipolar interactions, and these are known to play a significant role in the quantum dynamics of SMMs [13]; these are discussed below.

(ii) *The hyperfine couplings*: These have not been measured; however, we can estimate them based on experiments in other Mn-based systems. The most important are the local couplings of the Mn electron spins to their own nuclei. For the naturally occurring spin-5/2 $^{55}$Mn isotope, previous NMR measurements on different Mn$_{12}$-acetate systems [90] show hyperfine couplings of 220-230 MHz for the Mn$^{4+}$ ion, and two lines with frequencies of 280-290 MHz and $\sim 360$ MHz for the Mn$^{3+}$ ion. For the Mn$^{2+}$ ion a rather wider range of values has been found [92], ranging from $\sim 275$ MHz for Mn$^{2+}$ ions in PbF$_2$, to as high as $\sim 670$ MHz for Mn$^{2+}$ ions in simple MnF$_2$. We do not expect the couplings to be radically different in the Mn$_{12}$ wheel; thus for the Mn$^{2+}$ and Mn$^{3+}$ ions we will assume $A_{j,k=j} \sim 20$ mK and 15 mK respectively, with hyperfine lines spread across an energy range of 100 mK and 75 mK respectively.

The couplings of the Mn spins to the other nuclei will be predominantly dipolar. The hyperfine coupling to the protons is then easily found to range from $\sim$2.5 mK downwards, with most couplings in the range 0.5-1 mK. The coupling to N and any O or C nuclei is comparable.

(iii) *The magnetoelastic couplings*: To our knowledge, neither the anisotropy constants nor the sound velocity have been measured for this system. However it is easy to see that the corrections caused by the magnetoelastic interactions to the primary couplings are very small. Thus, e.g., if we take $|d_j| \sim 1$ K in (18), we find a correction $|\delta J_{ij}^{\alpha\beta\gamma\delta}| \sim 10^{-3}$ K or less, for typical values of the parameters in (20). Thus henceforth we will ignore the effect of magnetoacoustic interactions, even in the presence of disorder.

### II.B.2: Disorder

Any disorder in the system can have non-trivial effects: depending on the type of disorder, local superexchange and spin anisotropy couplings will be changed, and, because disorder typically breaks inversion symmetry, the DM interactions will be modified. The DM vector $\mathbf{D}_{12}$ no longer vanishes in the dimer case, and the four DM vectors in the tetramer case (see below) may be slightly altered (and in general all four of them will now be independent). There is a smaller effect on the magnetoacoustic interaction, and the effect on the hyperfine interactions is quite negligible.

Disorder effects are typically difficult to discuss quantitatively because there are many possible sources. The most obvious is site disorder, caused by defects, dislocations, or impurities in the crystal. This is present in all solids. In SMMs it can be particularly complex, because disorder in any of the ligand or lattice solvent groups surrounding the molecules can significantly disrupt the spin Hamiltonian of the central core. A well documented (and well characterized) case is that of Mn$_{12}$-acetate [64, 68]. In this case the disorder is discrete, i.e., one of a finite number of configurations is possible. However point or line disorder from defects will give a continuous distribution of perturbed spin Hamiltonians. If the spread in spin anisotropy parameters caused by defects is not small, it will completely obscure any structure coming from discrete disorder. The effect of disorder on the magnetoacoustic interaction is a little more subtle. Even for the simplest point disorder perturbation acting on the phonons, of strength $c_0$ at site $j$, the unperturbed spin-phonon coupling in (16) is replaced by

$$v(\mathbf{s}_j, \phi_{\mathbf{q}}) = -\sum_{\alpha\beta\gamma\delta}[c_0\delta^{\alpha\beta} + \mathbb{A}_j^{\alpha\beta\gamma\delta}s_j^\gamma s_j^\delta]\,u_j^{\alpha\beta}, \quad (21)$$

and thus generates, to lowest order, a perturbation $\delta k_j^{\alpha\beta} \sim O(\lambda)$ on the anisotropy constants. However the point disorder will also modify the spin orientations in unpredictable ways; this modifies not only the $k_{ij}^{\alpha\beta}$ but all of the other parameters of the spin Hamiltonian.

The situation is thus rather complex, so in what follows we will simply assume that disorder leads to a distribution in the values of the $k_{ij}^{\alpha\beta}$, $J_{ij}^{\alpha\beta}$ and the $D_{ij}^\alpha$, about their undistorted values. In the same way, the effect of disorder on the dipolar coupling will be modeled by adding a random $\delta m_j^\alpha$ to the unperturbed moments $m_j^\alpha$ in (13). Note that there is no reason to assume that these random distributions are symmetric; indeed in other systems they have been found to be highly asymmetric [67]. In this paper we will not be doing any explicit calculations of these disorder effects, so we do not specify the distributions explicitly.

The combination of disorder and a finite applied field also indirectly affects the system. We deal with this problem below, in the context of the dimer and tetramer models.

## III: DIMER AND TETRAMER MODELS

As we shall see in section IV, experiments indicate that a 'dimer' picture of the spin structure is a good starting point for analysis of the spin structure of the Mn$_{12}$ wheel. However the uncertainty in the numerical results for the $J_{ij}^{\alpha\beta}$, and the peculiar role played by the DM interaction, make it useful to extend this model, and so here we also develop a 'tetramer' model of the spin structure.

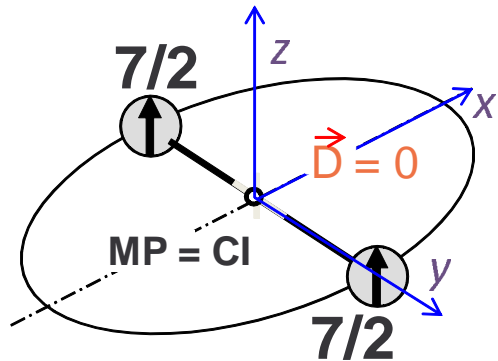

FIG. 3: (Color on-line) The dimer model for the Mn$_{12}$ wheel molecule, shown here for the inversion-symmetric case. The principal terms in this model are given in eqns. (22)-(24). The mid-point (MP) between the two magnetic units in the dimer coincides with the position of the center of inversion (CI), which forbids an antisymmetric DM interaction.

### III.A: Dimer model

The numerical results discussed earlier indicate that the Mn$_{12}$ molecule has a structure in which the superexchange interaction $J_6^0$ is substantially weaker than the others [77]. Experiments [70] confirm this. They also show that, to first approximation, one can model each half of the molecule as a single magnetic unit with spin $s = 7/2$; these two halves are then coupled *ferromagnetically* to give a total spin $S = 7$. Interestingly, the weakest coupling between two adjacent ions within the wheel (i.e. $J_6^0$), which allows the dimer description of the molecule, has to be *antiferromagnetic* to stabilize the *ferromagnetic* $S = 7$ dimer ground state (see Fig. 2). This is indeed borne out by the numerical calculations. Our dimer ansatz, along with the arrangement of the Mn spins it implies, are shown in Figs. 2 and 3.

If we follow the experiments and adopt the dimer approximation, then we can describe the system in terms of a Hamiltonian $\mathcal{H}_{SM}^{dim} = \mathcal{H}_0^{dim} + \mathcal{H}_1^{dim}$, where the bare term is

$$\begin{aligned}\mathcal{H}_0^{dim} = \ &[K^{\alpha\beta}(S_1^\alpha S_1^\beta + S_2^\alpha S_2^\beta) + O(S_{1,2}^4)] \\ &+ \mathcal{J}_{12}^{\alpha\beta}S_1^\alpha S_2^\beta - \mu_B(S_1^\alpha + S_2^\alpha)\mathbf{g}^{\alpha\beta}H_0^\beta, \quad (22)\end{aligned}$$

with $|S_l| = 7/2$, where $l = 1, 2$, and with again a simple quadratic biaxial form chosen for the anisotropy terms:

$$K^{\alpha\beta}S_l^\alpha S_l^\beta \sim -DS_{lz}^2 + E(S_{lx}^2 - S_{ly}^2). \quad (23)$$

We note that there is no DM term, of form $\mathcal{D}_{12}^\gamma S_1^\alpha S_2^\beta$, in $\mathcal{H}_0^{dim}$; it follows immediately from the inversion symmetry of the system that

$$\mathcal{D}_{12}^\gamma = 0. \quad (24)$$

We write the secondary terms for this dimer system as

$$\mathcal{H}_1^{dim} = U_{12}^{\alpha\beta} S_1^\alpha S_2^\beta + \delta\mathcal{J}_{12}^{\alpha\beta\gamma\delta} S_1^\alpha S_1^\beta S_2^\gamma S_2^\delta$$
$$+ \sum_k (\mathcal{A}_{1k}^{\alpha\beta} S_1^\alpha + \mathcal{A}_{2k}^{\alpha\beta} S_2^\beta) I_k^\beta, \qquad (25)$$

where we now have an interaction $U_{12}^{\alpha\beta}$ which is just the sum over all of the individual dipolar interactions between the two dimer halves (and notice that by making the dimer *ansatz*, we have implicitly dropped all internal dipolar interactions between electronic spins inside each dimer half−these are assumed to be absorbed into $\mathcal{H}_0^{dim}$). Thus we have

$$U_{12}^{\alpha\beta} S_1^\alpha S_2^\beta = \sum_{i=1}^{6} \sum_{j=7}^{12} V_{ij}^{\alpha\beta} s_i^\alpha s_j^\beta, \qquad (26)$$

where the $\mathbf{s}_i, \mathbf{s}_j$ are now assumed to be oriented according to the dimer *ansatz*, as shown in the Figure. In the same way as previously, we write this interaction in the form

$$U_{12}^{\alpha\beta} = U_{12}^0 \mathbb{D}_{12}^{\alpha\beta} \qquad (27)$$

where the tensor $\mathbb{D}_{12}^{\alpha\beta}$ has a similar (but not identical) dependence on the mutual orientation of the two dimer halves to the form in (15), and now $U_{12}^0$ defines the energy scale of the interaction.

In the same way, the new hyperfine interaction is given by summing over the individual couplings to each electronic spin in a given dimer half:

$$\mathcal{A}_{lk}^{\alpha\beta} S_l^\alpha I_k^\beta = \sum_{j \in l} A_{jk}^{\alpha\beta} s_j^\alpha I_k^\beta \qquad (l = 1, 2). \qquad (28)$$

There will in principle be corrections to the intra-dimer exchange interaction induced by magnetoacoustic couplings, of form $\delta\mathcal{J}_{12}^{\alpha\beta\gamma\delta}$, given likewise by summing over all pairs in each half:

$$\delta\mathcal{J}_{12}^{\alpha\beta\gamma\delta} S_1^\alpha S_1^\beta S_2^\gamma S_2^\delta = \sum_{i=1}^{6} \sum_{j=7}^{12} \delta J_{ij}^{\alpha\beta\gamma\delta} s_i^\alpha s_i^\beta s_j^\gamma s_j^\delta. \qquad (29)$$

As noted above, this correction is negligible, and we will ignore it.

Consider now the values of these secondary couplings. The dipolar coupling $U_{12}$ is dominated by the two individual couplings $V_{1,12}$ and $V_{6,7}$, between the two pairs of spins which join the two dimer halves. The distance between these pairs of spins is 3.473Å, and the two terms add. If we calculate the interaction in a point-dipole approximation, then we find that

$$U_{12}^0 \sim 0.049 \text{ K}, \qquad (30)$$

with corrections $\sim 0.01$ K from all the other dipolar interactions inside each dimer. This interaction is larger than it seems; we note that $U_{12}^{\alpha\beta} S_1^\alpha S_2^\beta$ varies over an energy range $\sim 1.8$ K, depending on the mutual orientation of the two dimer halves. When we come to compare with experiments on the $Mn_{12}$ wheels, we will see that although this dipolar interaction is probably considerably smaller than $\mathcal{J}_{12}$ above, it is not negligible.

Each dimer half couples to 6 Mn nuclear spins via the $\mathcal{A}_{lk}^{\alpha\beta}$. This means that even if we ignore the other nuclear spins, we deal with a manifold of $5^6 \sim 1.5 \times 10^4$ Mn nuclear spin states coupling to each dimer spin state. Using the numbers previously given, we find that these states are spread over an energy $\sim 0.53$ K. However, as is always the case with a spin bath [51], the density of bath states for each dimer half (here a multinomial distribution) is bell-shaped, with a much narrower linewidth. One then finds

$$E_0 \sim 105 \text{ mK}. \qquad (31)$$

Finally, we consider the effects of a field and disorder on the dimer system. The application of a finite transverse field has a dramatic effect. It is actually very interesting to consider this problem analytically, for a general set of parameters, but this turns out to be quite lengthy. The main objective of this paper is to see how this works out for the $Mn_{12}$ wheel system. Thus, in section IV below, we will simply derive the effects of a transverse field numerically, for the parameter values that are revealed by experiment.

We will treat disorder in this dimer system in the same way as described above, assuming that it generates a set of random components $\delta\mathcal{J}_{12}^{\alpha\beta}, \delta K^{\alpha\beta}$, to be added to the bare couplings; a set of parameters, $\delta M_l^\alpha$ to be added to the bare spin moments for each dimer half; and a set of random DM vectors $\delta\mathcal{D}_{12}^\alpha$. In all cases the mean value of these deviations is assumed to be zero.

There is also a dipolar field on each dimer from all the other dimers. In principle one can enlarge the description of the system to include these couplings in a fully quantum mechanical way. However (see section IV), we will treat these interactions in terms of a slowly-varying (in time and space) classical demagnetization field of order 0.01 T, corresponding to a Zeeman coupling of $\sim 0.04$ K to each spin-7/2 dimer half.

### III.B: Tetramer Model

There are a number of reasons for studying a tetramer model in the context of spin tunneling. In the case of $Mn_{12}$ itself, the large uncertainty as to the correct underlying spin Hamiltonian, and the debate about the effect of secondary interactions such as DM or dipolar couplings, makes this a useful avenue to explore. Moreover, the existing calculations of the superexchange parameters [77, 87] are not inconsistent with a tetramer *ansatz*. However there are also more general reasons for such an exploration: as we will see, some of the results (partic-ularly for the character of spin phase oscillations) are rather intriguing, and of quite general interest.

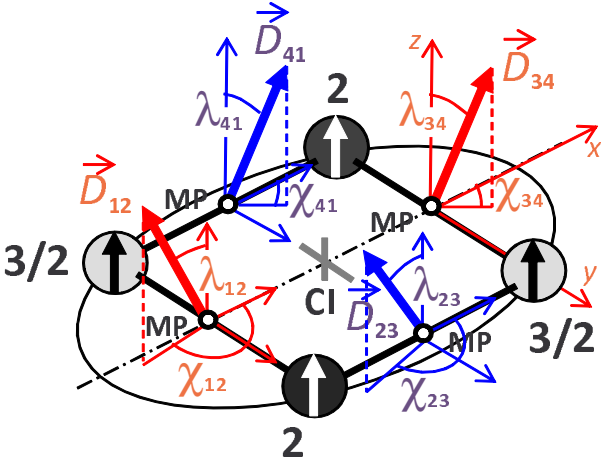

FIG. 4: (Color on-line) The tetramer model defined in the text (see eqns. (32)-(37)); the real space axes $\hat{x}$ (the hard axis), $\hat{y}$ and $\hat{z}$ (the easy axis) are also shown. Since the mid points (MP) on the bonds joining each sub-unit of the tetramer do not coincide with the molecule's center of inversion (CI), local DM interactions are allowed. We show here the most general case in which there is no inversion symmetry (no CI), so that the four DM vectors $\vec{\mathcal{D}}_{\mu\nu}$ can point in arbitrary directions. The polar and azimuthal angles $\bar{\lambda}_{\mu\nu}$, $\bar{\chi}_{\mu\nu}$ of each arbitrarily oriented DM vector are also shown.

The tetramer *ansatz* assumes that we have not two coupled subunits in the spin Hamiltonian, but four. Thus we can assume a Hamiltonian of form $\mathcal{H} = \mathcal{H}_0^{tet} + \mathcal{H}_1^{tet}$, where the primary term is written in terms of the four tetramer spins $\bar{\mathbf{S}}_\mu$ as:

$$\mathcal{H}_0^{tet} = \sum_{\mu=1}^{4} [\bar{K}_\mu^{\alpha\beta} \bar{S}_\mu^\alpha \bar{S}_\mu^\beta + O(\bar{S}_\mu^4)] - \mu_B \bar{S}_\mu^\alpha \bar{\mathbf{g}}_\mu^{\alpha\beta} H_0^\beta$$
$$+ \sum_{\mu<\nu} [\bar{\mathcal{J}}_{\mu\nu}^{\alpha\beta} + \epsilon_{\gamma\alpha\beta} \mathcal{D}_{\mu\nu}^\gamma] \bar{S}_\mu^\alpha \bar{S}_\nu^\beta, \qquad (32)$$

and where we again choose a simple quadratic biaxial form for the anisotropy

$$\bar{K}_\mu^{\alpha\beta} \bar{S}_\mu^\alpha \bar{S}_\mu^\beta \sim -\bar{D}_\mu \bar{S}_{\mu z}^2 + \bar{E}_\mu (\bar{S}_{\mu x}^2 - \bar{S}_{\mu y}^2). \qquad (33)$$

The way we will implement this tetramer scheme for the specific case of the $Mn_{12}$ wheel is shown in Fig. 4. Each dimer half is split into 2 sub-groups, each with a mixture of $Mn^{3+}$ and $Mn^{2+}$ spins (three total—see Fig. 2). Thus we have 4 sub-units in all, with total spin

$$\bar{S}_1 = \bar{S}_3 = 3/2; \qquad \bar{S}_2 = \bar{S}_4 = 2. \qquad (34)$$

The two sub-units inside a dimer half (e.g. $\bar{S}_1$ & $\bar{S}_2$) are then coupled ferromagnetically via the superexchange coupling $J_3^0$; in the limit of very large $J_3^0$ this will give a total ground state spin for the dimer halves of $S_1 = S_2 = 7/2$, but for any finite $J_3^0$ this state will mix with other states. Note that $J_3^0$ is actually found to be fairly small in some LDF calculations [87]. We re-emphasize that we do not believe there to be any positive evidence, so far, for this tetramer picture of the $Mn_{12}$ wheel. However there is also no evidence against it, and it is an interesting model to explore in its own right, particularly given the apparent weakness of $J_3^0$. Most importantly, the tetramer description represents the simplest extension of the dimer case that allows inclusion of DM terms while still respecting the inversion symmetry of the molecule. Indeed, the tetramer may also be viewed as a dimer, at low energies, when $J_3^0 >> J_6^3$.

If we assume an inversion symmetry for the molecule, then there are only two independent parameters in each of the terms in $\mathcal{H}_1^{tet}$, i.e.

$$\bar{\mathcal{J}}_{12}^{\alpha\beta} = \bar{\mathcal{J}}_{34}^{\alpha\beta} \quad \& \quad \bar{\mathcal{J}}_{23}^{\alpha\beta} = \bar{\mathcal{J}}_{14}^{\alpha\beta}$$
$$\bar{D}_1 = \bar{D}_3, \quad \bar{D}_2 = \bar{D}_4 \quad \& \quad \bar{E}_1 = \bar{E}_3, \quad \bar{E}_2 = \bar{E}_4. \qquad (35)$$

We note that for the $Mn_{12}$ wheel, if we ignore 2nd nearest-neighbor superexchange couplings, then $\bar{\mathcal{J}}_{12}^{\alpha\beta} = \bar{\mathcal{J}}_{34}^{\alpha\beta} \to J_6^0$, and $\bar{\mathcal{J}}_{23}^{\alpha\beta} = \bar{\mathcal{J}}_{14}^{\alpha\beta} \to J_3^0$.

There are similar constraints on the DM terms, although these are no longer ruled out by the inversion symmetry of the $Mn_{12}$ wheel. This symmetry requires that

$$\epsilon_{\gamma\alpha\beta} \bar{\mathcal{D}}_{12}^\gamma \bar{S}_2^\alpha \bar{S}_3^\beta = -\epsilon_{\gamma\alpha\beta} \bar{\mathcal{D}}_{34}^\gamma \bar{S}_4^\alpha \bar{S}_2^\beta$$
$$\epsilon_{\gamma\alpha\beta} \bar{\mathcal{D}}_{23}^\gamma \bar{S}_2^\alpha \bar{S}_3^\beta = -\epsilon_{\gamma\alpha\beta} \bar{\mathcal{D}}_{41}^\gamma \bar{S}_4^\alpha \bar{S}_1^\beta$$
$$\equiv +\epsilon_{\gamma\alpha\beta} \bar{\mathcal{D}}_{14}^\gamma \bar{S}_1^\alpha \bar{S}_4^\beta, \qquad (36)$$

and if the alignment of the spins $\bar{\mathbf{S}}_\mu$ also obeys the inversion symmetry, we have

$$\bar{\mathcal{D}}_{12} = \bar{\mathcal{D}}_{34}, \qquad \bar{\mathcal{D}}_{23} = \bar{\mathcal{D}}_{41}, \qquad (37)$$

i.e., two independent DM vectors.

In the next section we will explore some features of this model, allowing the exchange and anisotropy couplings, and the DM vectors, to be free parameters, whilst still satisfying the constraints (35) and (36).

The secondary interaction terms for this tetramer system are written as

$$\mathcal{H}_1^{tet} = \sum_{\mu<\nu} [\bar{U}_{\mu\nu}^{\alpha\beta} \bar{S}_\mu^\alpha \bar{S}_\nu^\beta + \delta \bar{\mathcal{J}}_{\mu\nu}^{\alpha\beta\gamma\delta} \bar{S}_\mu^\alpha \bar{S}_\nu^\beta \bar{S}_\nu^\gamma \bar{S}_\nu^\delta]$$
$$+ \sum_{\mu k} \mathcal{A}_{\mu k}^{\alpha\beta} \bar{S}_\mu^\alpha I_k^\beta, \qquad (38)$$

where the definition of the interaction parameters is an obvious generalization of what we did for the dimer problem above. Using the same kind of arguments as above (and noting that the distances between the relevant spins are 3.169 Å and 3.473 Å), we find that, in a point-dipole approximation, the couplings between each tetramer block coming from the nearest-neighbor dipolar interaction, are

$$\bar{U}_{12}^0 = \bar{U}_{34}^0 \sim 0.13 \text{ K} \quad \& \quad \bar{U}_{23}^0 = \bar{U}_{14}^0 \sim 0.10 \text{ K}, \qquad (39)$$

with corrections $\sim 25$ mK from the other dipolar interactions inside the tetramer. The dipolar couplings $\bar{U}_{13}^0, \bar{U}_{24}^0$ between opposite blocks of the tetramer are $< 5$ mK, so we neglect them. The dipolar couplings in (39) are not negligible, ranging over $\sim 1.2$ K and $\sim 0.9$ K respectively.

For the hyperfine couplings we have that each unit of the tetramer couples to 125 Mn nuclear spin states, spread over an energy range of $\sim 0.23$ K. One now finds a half-width

$$\bar{E}_0 \sim 60 \text{ mK}. \tag{40}$$

Again, we ignore magnetoacoustic corrections.

External fields and disorder are of course important for the tetramer model. We will study the influence of external fields numerically below, and disorder will be handled as for the dimer case, using extra random couplings whose mean is zero.

## IV: EXPERIMENTAL RESULTS

At the present time, experiments on SMMs always involve large numbers of oriented molecules in a single crystal. This necessitates consideration of the interactions between the molecules, notably the dipolar interaction. We do this briefly here, and then review the salient features of the tunneling relaxation experiments and their interpretation. We focus on the tunneling relaxation rates and spin phase oscillations associated with this tunneling.

### IV.A: Role of intermolecular Interactions

In the usual theory of tunneling relaxation for a large number of tunneling molecules [20], the effect of intermolecular dipolar interactions is treated using a BBGKY theory, in which the lowest-order pairwise interactions yield a field which is the sum of a molecular field (the demagnetisation field), which in general varies around the sample, and a fluctuation term. It is natural to ask whether one can use the same approach here, when we have a weak coupling between dimer halves of a molecule (or possibly between tetramer pieces). Clearly the results will depend on the ratio between the inter-dimer exchange $J_{12}$ and the strength of the dipolar terms, of which there are two to consider; we have the inter-dimer dipolar interaction, of strength $U_o \sim 0.05$ K (see eqns. (26), (27), and (30)), and also the intermolecular dipolar interaction

$$V_{nm}^{dip} = \mathbf{U}_{nm} \mathcal{S}_n^\alpha \mathcal{S}_m^\beta \tag{41}$$

where we have labeled the molecular sites by $n, m, ...,$ and $\mathcal{S}_n$ is the molecular spin of the entire molecule on the $n^{\text{th}}$

lattice site (so that $|\mathcal{S}_n| = 7$). Writing, in the same way as (27), the interaction is

$$\mathbf{U}_{nm} = \mathbf{U}_{nm}^0 \mathbb{D}_{nm}^{\alpha\beta}. \tag{42}$$

If we assume a point-dipole approximation, then for a pair of nearest-neighbor molecules, $(U_{nm}^0 \sim 6 \times 10^{-4}$ K (recall the nearest-neighbor intermolecular distance is $\sim 16$ Å); this implies the typical range of energies for the nearest-neighbor interaction of $\sim 80$ mK, and up 2-3 times larger for a macroscopic sample (depending on whether the molecular spins are polarized or randomly oriented).

We see that the intermolecular dipole interactions are actually rather small, in comparison with the other interactions in the system. This allows us to make an important approximation, viz., we replace the entire dipolar interaction by a demagnetisation field, of typical magnitude $H_{dm} \sim 0.01$ T. The transverse part of this is very small compared to the transverse fields of interest in the experiments. The longitudinal part $H_{dm}^z$ will be added to any applied $H_z$, so that the total longitudinal energy bias on a spin $S$ will be $\xi_z = g\mu_B S_z(H_z + H_{dm}^z)$.

### IV.B: Quantum Relaxation Measurements

In a typical low temperature field sweep measurement on SMM systems, one measures the time-dependent magnetization, $M_z(t; H_z, \mathbf{H}_\perp)$, along the easy axis, as one sweeps the longitudinal field, $H_z(t)$. From the derivative, $dM_z/dH_z$, at a given sweep rate and transverse field, one tries to extract the tunneling transition amplitudes, $\Delta_{nn'}(\mathbf{H}_\perp)$, at the resonant transition fields (involving the levels $n$ & $n'$ of the tunneling spin system). The theoretical justification for this procedure is as follows.

(a) If the sweep rate is low, then the relaxation rate of the magnetization is given by a relaxation time, $\tau_Q(\mathbf{H}_\perp)$, which for a sample of arbitrary shape is given by [20]

$$\tau_Q^{-1}(\mathbf{H}_\perp) = \frac{\xi_0^2 N(\xi_z)}{W_0} \frac{2|\Delta_{nn'}(\mathbf{H}_\perp)|^2}{\pi^{1/2}\Gamma_2}. \tag{43}$$

Here, $N(\xi_z)$ is the normalized distribution of longitudinal bias energies, from fields acting on the molecules, as a function of the longitudinal bias, $\xi_z = g\mu_B S_z H_z$, coming from an applied field; $W_0$ is its width. The parameter $\xi_0$ is an effective energy range over which the nuclear spins can influence the tunneling, and $\Gamma_2$ the energy range over which the nuclear manifold fluctuates in energy via $T_2$ transitions. In most experiments one has $\xi_0 \sim \Gamma_2 \sim E_0$, where $E_0$, defined earlier, is the half-width of the nuclear density of states (for a more precise discussion, see refs. [13, 20, 93]). Thus, in slow sweep experiments, one has

$$\tau_Q^{-1}(\mathbf{H}_\perp) \sim \frac{E_0}{W_0} N(\xi_z)|\Delta_{nn'}(\mathbf{H}_\perp)|^2. \tag{44}$$

In the preceding sections we have estimated $E_0$; $W_0$ and $N(\xi_z)$ can in principle be extracted from the experimental hysteresis curves. Thus, one can get $|\Delta_{nn'}(\mathbf{H}_\perp)|^2$ from these measurements.

(b) For very high sweep rates, it has been quite common to use the simple Landau-Zener-Stuckelberg formula [94], adapted to a set of non-interacting molecules [95], according to which the transition rate at a bias $\xi_z$ is

$$\tau_{LZ}^{-1} = 1 - N(\xi_z)\frac{\pi|\Delta_{nn'}(\mathbf{H}_\perp)|^2}{2(d\xi_z/dt)}. \qquad (45)$$

This formula is valid in the limit of very fast sweeps. However its application to experiments is too naive, even for the quite fast sweeps used in some experiments. The effects of both nuclear spins [29, 32, 33, 35] and dipolar interactions [29, 31, 34, 35] persist to sweep rates well above those used in any experiments thus far, and give substantial corrections to (45). As noted in the introduction, there is no generally accepted formula for this 'intermediate sweep rate' regime. Most treatments include only one or other of the nuclear bath and the inter-molecular dipolar interactions (see however ref. [35] which does include both). We also note in passing that the treatment of the nuclear bath as a simple classical noise source is not in general valid [13].

What is not always recognized is that there must be a large intermediate range of sweep rates between the very fast limit, where we expect (45) to be valid, and the slow limit where we expect (44) to be valid. This is because the slow sweep result breaks down when the sweep rate begins to compete with the dynamics of 'hole burning' by the tunneling, and this depends on rather slow dipolar interaction processes [29, 31, 93, 96]. This implies that one is in the slow sweep regime provided the sweep rate $\dot{\xi}_z$ satisfies

$$\dot{\xi}_z \ll E_0 \tau_Q^{-1}(\mathbf{H}_\perp), \qquad (46)$$

which for the present case, using the values for $E_0$ found in section III (eqns. (31) and (40)), implies $\dot{\xi}_z \ll 10^{-3}$ Ts$^{-1}$. This is an order of magnitude estimate only, but it implies that in the experiments (where $\dot{\xi}_z = 0.2$ T/min), one may already already be out of the slow sweep regime.

However one does not reach the fast sweep limit until the time, $\Delta t \sim |\Delta_{nn'}|/\dot{\xi}_z$ (required for the field to sweep through a tunneling resonance), is much less than the time it takes the fluctuating nuclear bias field to sweep through the same resonance [29, 32, 33, 51]. This latter time is $\tau_{fl}(\Delta) \sim |\Delta_{nn'}|^2 T_2/E_0^2$, where $T_2$ is the transverse nuclear spin relaxation time [51]. Accordingly, we have

$$\dot{\xi}_z \gg \frac{E_0^2}{T_2|\Delta_{nn'}|} \qquad (47)$$

as the condition required to reach the fast sweep regime. For values of $|\Delta_{nn'}|$ in the range $10^{-6} - 10^{-5}$ K, and again using (31) and (40), we see that this requires sweep rates $> 10^3$ Ts$^{-1}$, even for rather low values of $T_2$.

We thus conclude that the experiments were done in the intermediate sweep rate regime, where no rigorous theory yet exists. For this reason the simplest possible method was employed in the data analysis: we used the unrenormalized Landau-Zener formula (45). Undoubtedly errors arise in this case, but we believe that they will give less than an order of magnitude error in the extracted values of $|\Delta_{nn'}|$.

Single molecule tunneling resonances show up in any experiment as peaks in $dM_z/dH_z$, coming from peaks in $N(\xi_z)$ centered around the SMM resonance transition fields. If one accepts the usual theoretical interpretation of these peaks, the lineshape comes from: (i) a combination of intermolecular fields which vary around the sample, and vary slowly in time; (ii) static disorder in the sample; and (iii) broadening caused by the nuclear spins.

From experiments one can extract the following information:

(i) The level-crossing fields and, hence, the level-crossing energies, allow certain deductions about the form of the underlying spin Hamiltonian.

(ii) The zero transverse field quantum relaxation rates, which give approximate values for the tunneling matrix elements, along with the values of the transverse fields corresponding to Berry phase minima, provide further information about the spin Hamiltonian.

(iii) The resonance lineshapes provide information about the distribution of dipolar and/or disorder fields (i.e. $g$-, $D$- and $E$-strain or tilts), though it can be challenging to deconvolute the various contributions when they are of similar magnitude.

The results for a slow sweep experiment (with $dH_z(t)/dt = 0.2$ T/min), on a single crystal of Mn$_{12}$ wheel molecules are shown in Fig. 5, from ref. [70]. The low-T resonances are rather broad, with linewidths ranging from $0.06 - 0.1$ T. These linewidths correspond to an energy spread $\sim 0.3 - 0.5$ K in the dimer picture (if $|\tilde{S}_z^i| = 7/2$). Such linewidths cannot be accounted for solely by intermolecular dipolar fields; note that the typical demagnetization field is estimated to be of order $0.01$ T. This could be checked by comparing results with initially annealed and initially field-cooled states. Most likely, the longitudinal field linewidths are due mainly to disorder, i.e., there are significant, at least partially random, longitudinal interactions caused primarily by strains in $D$ and $J$. The same arguments apply to the tetramer *ansatz*, as we will discuss below.

Consider now the behavior of the transition rates as a function of transverse field (Fig. 6). We see 'Berry phase' minima. However, the positions of these minima clearly depend on which resonance we are looking at. The minima are strongly smeared (so that nowhere is the maximum amplitude extracted for $\Delta(\mathbf{H}_\perp)$ more than 4 times the minimum). From the smearing of the $k =$ $1(S)$ resonance, one deduces that the field spread in the transverse direction is $\sim 0.08\ T$. This is roughly what was found for the longitudinal spread, indicating that the disorder is roughly isotropic in this experiment.

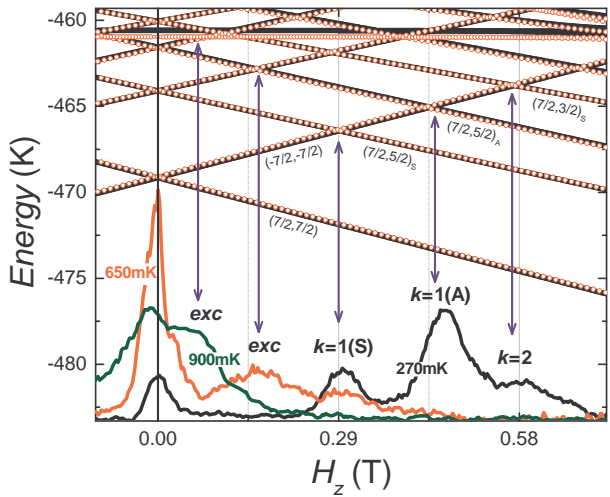

FIG. 5: (Color on-line) Measured derivative $dM_z/dH_z$ as a function of longitudinal field $H_z$, for a single crystal of $Mn_{12}$ wheel molecules. The data were taken at a sweep rate $dH_z/dt = 0.2$ T/min., for a variety of temperatures (from ref. [70]). We also show the energies of the lowest eigenstates in the dimer (straight lines) and tetramer models (open circles), calculated using the parameters in the text. The various resonances which arise at level crossings are also identified in this Figure.

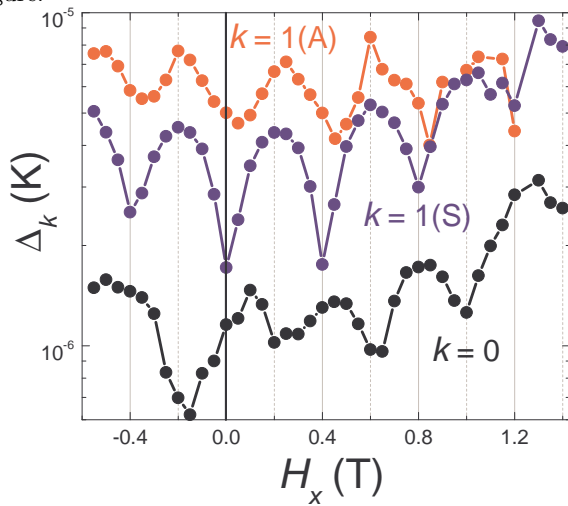

FIG. 6: (Color on-line) Tunnel splittings, extracted from field sweep experiments, for a single crystal of $Mn_{12}$ wheel molecules (from [70]). A simple Landau-Zener formula was used to extract these results.

## V: ANALYSIS OF THE DIMER AND TETRAMER MODELS

We now explore some of the consequences of the dimer and tetramer models for the $Mn_{12}$ wheel system, concentrating particularly on the role of the secondary interac-tions, and on the role of the DM interactions once any inversion symmetry-breaking is introduced.

### V.A: Dimer model Analysis

The fitting of the experimental data to a dimer model was already discussed in our previous work [70]. Here we revisit this question, examining the role of disorder, dipolar fields, and possible symmetry-breaking in the system.

#### V.A.1: Fits to the dimer model

In our previous work on the $Mn_{12}$ wheel [70], we analyzed the data using two forms for the dimer model. First, we used the simple dimer *ansatz* Hamiltonian $\mathcal{H}_0^{dim}$ of eqns. (22)-(24), using the following parameters:

$$D = 0.865\ K \qquad E = 0.156\ K$$
$$\mathcal{J}_{12} = 0.39\ K. \qquad (48)$$

These parameters are arrived at by numerical fits against both the level crossing longitudinal fields (Fig. 5) and the period of the Berry phase oscillations for the symmetric $k = 0$ and $k = 1(S)$ resonances (Fig. 6). Note that the reason why we see spin phase oscillations so clearly in this system is precisely because the numbers in (48) are not large; the transverse field Zeeman coupling then easily competes with the spin anisotropy. Thus, quite small perturbations can in principle seriously affect this dimer system, and we have to consider the effect of the secondary interactions.

Our previous work also considered, in its Appendix [70], a second form for the dimer model, which added a small DM vector $\mathcal{D}_{12}$ between the dimer halves, i.e., which added a term

$$\delta\mathcal{H} = \mathcal{D}_{12} \cdot \mathbf{S}_1 \times \mathbf{S}_2 \equiv \epsilon_{\gamma\alpha\beta}\mathcal{D}_{12}^{\gamma}S_1^{\alpha}S_2^{\beta}, \qquad (49)$$

where, as before, we assume $|\mathbf{S}_{1,2}| = 7/2$.

The existence of such a term in this dimer model automatically implies that the inversion symmetry of each molecule is broken. This assumption was made to give one possible explanation for the existence of the transitions between symmetric and antisymmetric states, that were seen at the antisymmetric resonance transition $k = 1(A)$. Indeed, one can explain most features of the observed transverse field behavior of the tunneling amplitude $\Delta_{k=1(A)}$ of the antisymmetric $k = 1(A)$ resonance with the use of a finite DM vector appropriately directed in space, as we show below.

Given the assumed symmetry breaking, one can estimate the size of DM vector that might be required simply from what we know of the spin-orbit coupling of typical Mn ions. This was discussed in section II.A.2; adapting this discussion here, we see that we we expect $|\mathcal{D}_{12}| = 2 \times d_e/4S_1S_2 = 2d_e/49$, where as we previously noted, typical values of $d_e \sim 0.2 - 1\,K$ (the factor of 2 comes because there are 2 exchange links between each dimer half). Thus we might expect $|\mathcal{D}_{12}| \sim 8 - 40$ mK. In fact the fitting parameters that worked best to explain the experimental results used $\mathcal{J}_{12}, E$ and $D$ as in (48) above, but added a DM vector with magnitude

$$\mathcal{D}_{12} = 34 \text{ mK} \qquad (\sin \theta_{12} = |\mathcal{D}_{12}/\mathcal{J}_{12}| = 0.085), \quad (50)$$

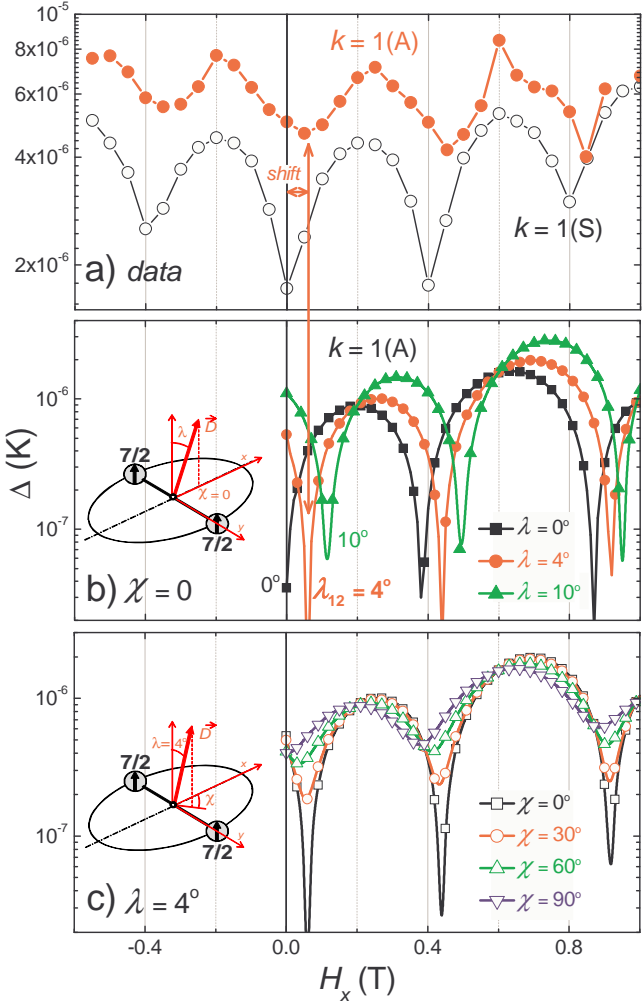

FIG. 7: (Color on-line) (a) The tunneling amplitudes for the $k = 1(S)$ and $k = 1(A)$ resonances, extracted from experiment (see also Fig. 6); (b) The tunneling amplitude $\Delta_{k=1(A)}$ for resonance $k = 1(A)$, calculated for the dimer model using the parameters in eqn. (48), and with the DM vector, $\hat{\mathbf{d}}_{12}$, defined in (51), tilted an angle $\lambda$ away from the $z$-axis in the $zx$-plane (i.e., $\chi = 0$); (c) The amplitude $\Delta_{k=1(A)}$ calculated with the same interaction parameters (eqn. (48)), now using $\lambda = 4°$, and for various rotation angles $\chi$ of $\hat{\mathbf{d}}_{12}$ out of the $zx$-plane.

(i.e., a spin-orbit angle $\theta_{12} = 5°$), and an orientation defined by a unit vector $\hat{\mathbf{d}}_{12}$, such that

$$\mathcal{D}_{12}^{\gamma} = \mathcal{D}_{12}\hat{\mathbf{d}}_{12}$$
$$\hat{\mathbf{d}}_{12} = (\sin \lambda \cos \chi, \sin \lambda \sin \chi, \cos \lambda), \quad (51)$$

with the director $\hat{\mathbf{d}}_{12}$ defined by the Bloch sphere angles $\lambda, \chi$.

Fig 7(a) directly compares the symmetric $k = 1(S)$ and antisymmetric $k = 1(A)$ resonance data. There are two crucial points to note: (i) the locations of the $k = 1(A)$ minima are shifted relative to $k = 1(S)$ such that the former are *not* symmetric with respect to the applied transverse field; and (ii) the minima are much more rounded for $k = 1(A)$ compared to $k = 1(S)$. We note that exactly the same behavior has been observed by Wernsdorfer *et al.* in Ref. [71]. These two observations highlight the very different character of these two resonances. In particular, (i) hints at the antisymmetric nature of $k = 1(A)$, though one must be careful because reversal of $H_\perp$ is not equivalent to an inversion operation (not to mention the fact that a fixed longitudinal field is applied in these experiments). Furthermore, (ii) cannot be explained simply on the basis of random dipolar fields, or by $D$ strain, which would be expected to influence both resonances in the same way.

Fig. 7(b) and (c) display the dimer model predictions for the behavior of $\Delta_{k=1(A)}$ as a function of a transverse field, $H_x$, applied along the hard $x$-axis, and for various orientations of the DM vector. Fig. 7(b) shows how a tilt of $\mathbf{d}_{12}$ away from the easy $z$-axis, in the $z$-$x$ plane, produces a rapid shift of the Berry phase minima with respect to zero transverse field, as observed in the data (this was first noted in Ref. [71]). A tilt $\lambda = 4°$ already reproduces the shift of 0.07 T observed for $k = 1(A)$ experimentally, though it fails to account for the rounding of the minima. It is then interesting to look at the effect of a rotation of $\hat{\mathbf{d}}_{12}$ out of the $z$-$x$ plane (so that $\chi \neq 0$). Fig. 7(c) shows the result: interestingly, the results are consistent with the observed rounding of the Berry phase minima at resonance $k = 1(A)$ when $\chi \sim 30°$. We comment more on this in section IV.C. where we analyze the experiments in the context of the tetramer model.

*V.A.2: DM vector plus external field*

It is well known that, in conventional Berry phase oscillations for a single 'giant spin' with biaxial (easy axis, easy plane) symmetry, any rotation of the transverse (in-plane) field $\mathbf{H}_\perp$ away from the hard $\hat{x}$-axis will destroy the Berry phase zeroes (see, e.g., ref. [11]). The simplest non-perturbative way to understand this is to note that the rotation pulls the 2 semiclassical tunneling paths away from their symmetric disposition on either side of the hard axis; one of the paths on the Bloch sphere becomes shorter, and moves through a region of lower potential, whereas the other actually becomes longer and moves through a region of higher potential. Thus, very rapidly, a large difference appears between the action of the two paths as a function of the angle $\phi_0$ between $\mathbf{H}_\perp$ and the hard axis, and the usual pattern of Berry phase

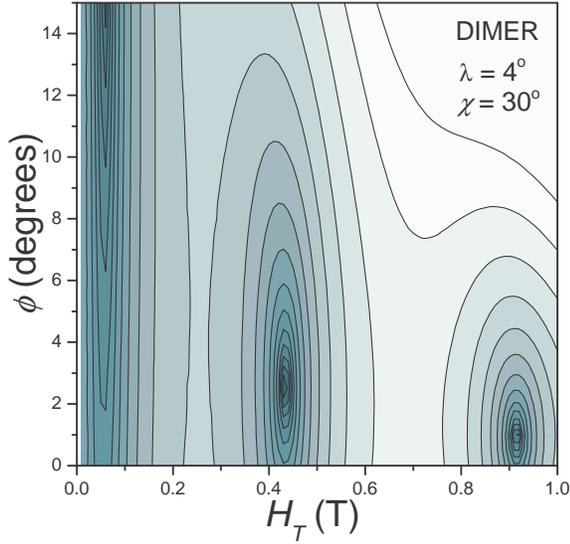

FIG. 8: (Color on-line) Contour plot of the calculated tunnel splitting for the antisymmetric $k = 1(A)$ transition, in the dimer model, shown as a function of the strength and direction of a transverse field $\mathbf{H}_\perp$, oriented at an angle $\phi_\circ$ with respect to the hard $x$-axis. We again assume the parameters given in eqns. (48) & (50), along with a DM vector oriented so that $\lambda = 4°$ and $\chi = 30°$.

zeroes is rapidly eliminated as the spin preferentially follows one of the 2 paths.

What we wish to point out now is that the effect of this field rotation in the dimer tunneling model is quite different. There is a very interesting interplay between the effect of a finite DM in-plane vector $\mathcal{D}_{12}$ and the in-plane applied field $\mathbf{H}_\perp$. In fact, one can *compensate* the other. To see this, we show in Fig. 8 the amplitude of the tunnel splitting for the dimer model, including both a finite DM vector and a rotated transverse field. To be specific, we have assumed that the DM vector $\mathcal{D}_{12}$, with director $\mathbf{d}_{12}$ defined by the Bloch sphere angles $\lambda, \chi$, takes the value

$$(\lambda, \chi) = (4°, 30°)$$
$$\mathcal{D}_{12} = 34 \text{ mK}, \tag{52}$$

and we then vary both the strength of the transverse field, $H_\perp$, and its orientation, $\phi_\circ$, relative to the hard axis. The result is a rather beautiful picture of the evolution of the tunneling amplitude, with Berry phase zeroes at various discrete points in $(H_\perp, \phi_\circ)$ space. Although we do not show it here, one may also examine the evolution of these Berry phase zeroes as a function of the DM vector strength and orientation, as well as the strength and orientation of the applied field. The results are complex and fascinating, and could be tested in experiment.

## V.B: Tetramer model fits

As discussed earlier in the paper, it is useful to explore a tetramer model, both for its intrinsic interest, and because of the uncertainty about the real values of the couplings in the $Mn_{12}$ wheel molecule. Not surprisingly, the tetramer model possesses many additional parameters: even in the perfectly inversion-symmetric molecule, one has two independent sets of DM vectors, exchange couplings, and anisotropy constants. Thus, in what follows, we will merely hint at the richness of the model, with results presented for a few interesting special cases.

The model was defined in eqns. (32) & (33); when inversion symmetry is obeyed, one also requires the constraints (35)-(37). In all of what follows we will further simplify things by making the (unrealistic) assumption that the anisotropy parameters are the same for all four tetramer sub-units. Thus, we fix

$$\bar{D}_\mu = \bar{D} = 1.65 \text{ K} \quad (\mu = 1, ..4)$$
$$\& \quad \bar{E}_\mu = \bar{E} = 0.36 \text{ K} \quad " \quad ". \tag{53}$$

We also fix the superexchange constants such that

$$\bar{\mathcal{J}}_{12}^{\alpha\beta} = \bar{\mathcal{J}}_{34}^{\alpha\beta} \to \bar{\mathcal{J}}_w = 0.74 \text{ K}$$
$$\& \quad \bar{\mathcal{J}}_{23}^{\alpha\beta} = \bar{\mathcal{J}}_{14}^{\alpha\beta} \to \bar{\mathcal{J}}_S = 74 \text{ K} \equiv 100 \ J_w. \tag{54}$$

Here, the subscripts $S$ and $w$ refer, respectively, to the strong and weak bonds. The values of $\bar{\mathcal{J}}_w$, $\bar{D}$ & $\bar{E}$ are chosen to fit the experimental data. The value of $J_S$ is chosen somewhat arbitrarily: it in fact represents a strong-coupling limit which partially (but, as we will see, not completely) mimics the dimer behavior.

Thus, the free parameters in the model are now the 4 DM vectors. We write the components of these as

$$\bar{\mathcal{D}}_{\mu\nu}^\gamma = \bar{\mathcal{D}}_{\mu\nu} \hat{d}_{\mu\nu}^\gamma$$
$$\hat{d}_{\mu\nu}^\gamma = (\sin\lambda_{\mu\nu}\cos\chi_{\mu\nu}, \sin\lambda_{\mu\nu}\sin\chi_{\mu\nu}, \cos\lambda_{\mu\nu}), \tag{55}$$

so that $\hat{d}_{\mu\nu}^\gamma$ is the unit vector in the direction of $\bar{\mathcal{D}}_{\mu\nu}^\gamma$, with polar and azimuthal angles $\lambda_{\mu\nu}, \chi_{\mu\nu}$, respectively. For what follows, we have fixed the magnitudes of the DM vectors as follows:

$$\bar{\mathcal{D}}_w \equiv \bar{\mathcal{D}}_{12} = \bar{\mathcal{D}}_{34} = 0.103 \text{ K}$$
$$\bar{\mathcal{D}}_S \equiv \bar{\mathcal{D}}_{23} = \bar{\mathcal{D}}_{41} = 10.3 \text{ K} \tag{56}$$

### V.B.1: Inversion symmetric tetramer

As discussed in III.B, the inversion symmetric system has only two independent DM vectors. Even so, this means that its properties depend in general on two different vector orientations, as well as their magnitudes. Rather than explore this 6-dimensional problem, we first

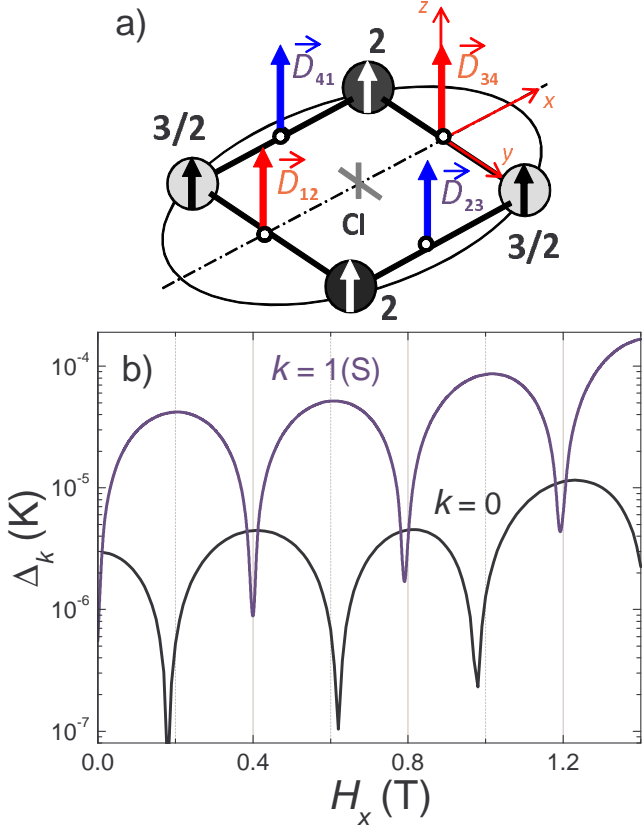

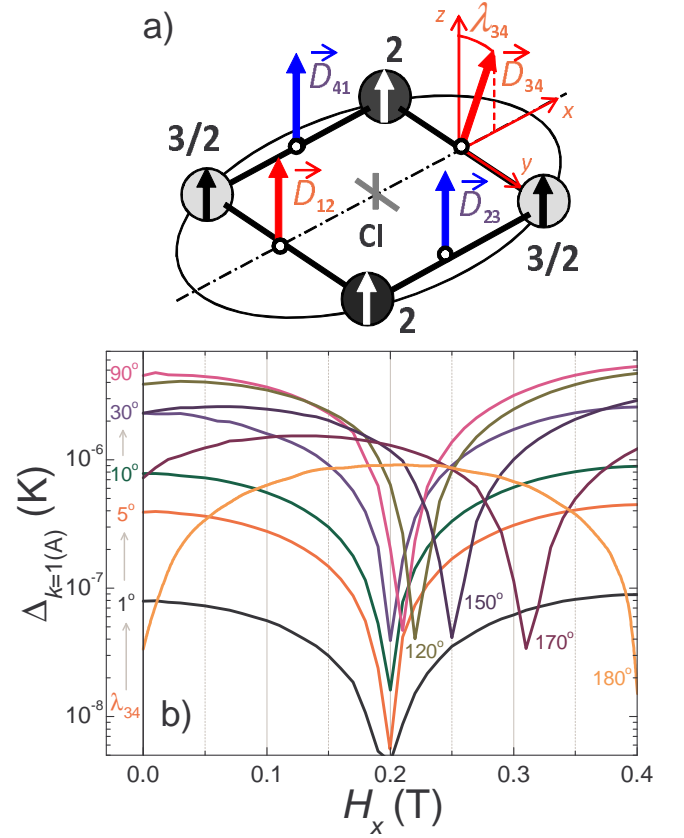

FIG. 9: (Color on-line) In (a) we show the arrangement of the DM vectors for the tetramer model with inversion symmetry specified in eqn. (57). (b) The calculated $k = 0$ and $k = 1(S)$ tunneling amplitudes for the inversion symmetric tetramer model. The parameters used are those specified in eqns. (53), (54) and (56), with the two DM vectors along $\hat{z}$ (cf. eqn. (57)), as shown in (a).

FIG. 10: (Color on-line) Calculated tunneling amplitude for the antisymmetric $k = 1(A)$ resonance, for the tetramer model with broken inversion symmetry, as a function of transverse field $H_x$ along the $\hat{x}$-axis. Here, the single DM vector, $\bar{\mathcal{D}}_{34}$, is tilted away from the others by an angle $\lambda_{34}$ in the $zx$-plane (compare eqns. (58) & (59)).

consider one particular case: we assume the parameter values in (56), and specify the directions as

$$\hat{\mathbf{d}}_w = \hat{\mathbf{d}}_S = \hat{z}, \qquad (57)$$

i.e., we assume that all of the DM vectors are parallel, so that $\lambda_{\mu\nu} = 0$ for all of the tetramer links. This situation is illustrated in Fig. 9(a). We note that the inversion symmetry requires parallel DM vectors of equal magnitude on opposite bonds of the $Mn_{12}$ wheel molecule.

As one might expect, the relaxational dynamics of this system in a transverse field is the same as what one finds for the dimer model. In Fig. 9(b) we show the calculated tunnel splitting for this system, for both of the symmetric ($k = 0$ and $k = 1(S)$) resonances. There is no $k = 1(A)$ resonance, because of the inversion symmetry. As can also be seen in this figure, the experimental data for the symmetric resonances agree well with the calculated curves, except for the absence of the antisymmetric resonance. This is hardly surprising−the system we have chosen, with $\mathcal{J}_S \gg \mathcal{J}_w$, and only two independent DM vectors, essentially mimics the inversion-symmetric dimer.

It is clear that no change in the values of $\bar{\mathcal{J}}_S$, $\bar{\mathcal{J}}_w$, $\theta_{\mu\nu}$, or in the DM angles $\hat{\mathbf{d}}_w$, $\hat{\mathbf{d}}_S$, will give any antisymmetric resonant tunneling. This is still forbidden by the inversion symmetry. However, changing these parameters will change the characteristics of the symmetric tunneling. In particular, lowering $\bar{\mathcal{J}}_S$ to a value more like $\bar{\mathcal{J}}_w$ (which turns the system into a genuine tetramer, with 4 partially decoupled sub-units), introduces a host of new symmetric resonances which are not seen experimentally in the $Mn_{12}$ wheel system. The behavior of these symmetric resonances becomes even more complex if we then allow the two DM vector directions to be independent. We do not pursue these possibilities here, for lack of space. However, one cannot completely rule out some combination of parameters that can account for the experiments entirely on the basis of symmetric tunneling transitions.

*V.B.2: Broken symmetry tetramers*

We now consider several different examples of tetramers in which the inversion symmetry is broken. In what follows, we emphasize that the only place we allow this symmetry-breaking to appear is in the directions of the DM vectors; their magnitudes are as given above.

(i) *Single tilt in the zx-plane*: As we have seen previously, if we respect the inversion symmetry, then any pair of DM vectors on superexchange bonds related by inversion *must* be equal. The only way to break the inversion symmetry by rotating DM vectors is to then let one of the vectors in such a pair rotate away from the other. Our first example of this kind makes the following assumptions:

$$\lambda_{12} = \lambda_{23} = \lambda_{41} = 0 \qquad (58)$$

$$\chi_{34} = 0, \qquad (59)$$

but the angle $\lambda_{34}$ is allowed to vary. In other words, we allow one of the four DM vectors to rotate away from the $\hat{z}$ axis, in the $zx$-plane, while keep the other three fixed along $\hat{z}$, as illustrated in Fig. 10(a).

The first obvious effect of this tilting of $\mathcal{D}_{34}$ away from the $\hat{z}$ axis is to induce an antisymmetric resonance. Fig. 10(b) shows the dependence of the $k = 1(A)$ tunnel splitting on $\lambda_{34}$ and $H_x$; its behavior turns out to be rather interesting. For zero transverse field ($H_x = 0$) and small $\lambda_{34}$ ($\ll 30°$), the splitting emerges linearly with $\sin(\lambda_{34})$, i.e. the projection of $\mathcal{D}_{34}$ onto the $xy$-plane. For finite $H_x$ and small $\lambda_{34}$, $\Delta_{k=1(A)}$ exhibits a Berry phase zero at the same transverse field as the symmetric $k = 0$ resonance (see Fig. 9(b)). As $\lambda_{34}$ increases further, the sharp increase in $\Delta_{k=1(A)}$ at $H_x = 0$ levels off, reaching a maximum when $\lambda_{34} \sim \pi/2$. Meanwhile, the position of the Berry phase minimum remains almost unchanged at $H_x \sim 0.2$ T for tilt angles up to $\lambda_{34} \sim \pi/2$. However, at right around $\lambda_{34} \sim \pi/2$, the position of the minimum begins to shift very quickly in transverse field, until it changes phase completely for $\lambda_{34} = \pi$, such that the pattern of Berry phase minima is the same as for the symmetric $k = 1$ resonance (see Fig. 9(b)). Clearly, small deviations from $\lambda_{34} = \pi$ will lead to the situation observed experimentally; we consider this case in more detail below.

In an effort to mimic possible effects due to disorder, it is desirable to see what happens when we perturb away from the results shown in Fig. 10. X-ray studies suggest that any perturbation of the inversion symmetry must be random; were this not the case, one would expect to observe clear signatures in the X-ray spectra, as was the case for $Mn_{12}$-acetate [64]. For this analysis, we maintained the secondary constraint

$$\lambda_{23} = \lambda_{41} = 0, \qquad (60)$$

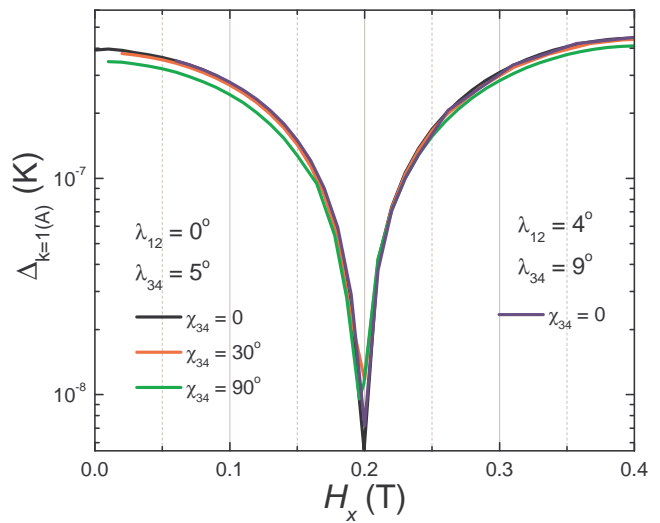

FIG. 11: (Color on-line) Tunneling amplitude, $\Delta_{k=1(A)}$, calculated as a function of a transverse field applied along the hard anisotropy $x-$axis, for the tetramer model in which $\lambda_{23} = \lambda_{41} = 0$, but where we allow small tilts of the other two DM vectors away from the $\hat{z}$ axis (small $\lambda_{12}\&\lambda_{34}$), as well as different finite values of $\chi$. This mimics the effects of possible molecular distortions, caused by disorder, which locally break the inversion symmetry.

then: (a) tested the effect of dropping the constraint in (59), allowing the azimuthal angle $\chi_{34}$ to rotate arbitrarily, while fixing $\lambda_{34} = 5°$; and (b) tested the effect of breaking the constraint in (58), allowing $\lambda_{12}$ to take small values. The results are shown in Fig. 11, for both cases. We note that while these deviations from perfect inversion symmetry may seem small, they likely represent colossal distortions of the molecules, i.e. changes in bond angles comparable to the employed variations in the relative orientations of the DM vectors ($\lambda_{12} - \lambda_{34}$). X-ray studies are not consistent with this degree of disorder. Nevertheless, we briefly summarize the results of this analysis.

Comparing Figs 10 and 11, we see that there is almost no change in the results shown above, even with arbitrary rotations away from the $zx$ plane. Thus, we conclude that the only effect of small deviations of the DM vectors from the parallel orientations required by inversion symmetry is to *switch on* the $k = 1(A)$ resonance; its pattern of Berry phase zeroes, meanwhile, does not shift noticeably from that of the $k = 0$ resonance. This conclusion is rather striking: it confirms that the effect of (strong) disorder on the DM vector orientations would be to create an antisymmetric resonance. For random disorder, the distribution of tunnel splittings will be smeared, leading to a smearing of the resonance. However, such disorder *does not* in any way reproduce several of the key results of the experiments, viz.: (1) the calculated Berry phase minimum is in completely the wrong location, i.e. well away from $H_T = 0$; and (2) there is no shift of the $k = 1(A)$ Berry phase pattern, reflecting the asymmetry observed in the experiments. To get such a shift, a very large perturbation is required, in which one or both of the DM vectors is rotated nearly 180° with respect to its opposite pair.

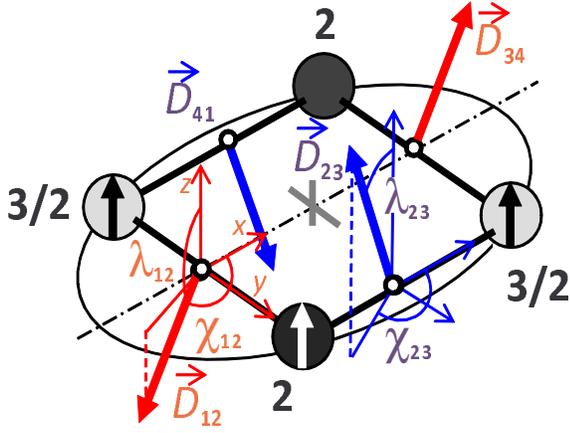

FIG. 12: (Color on-line) Orientation of the DM vectors for a tetramer model with broken inversion symmetry: the 'antiparallel' case defined by eqn. (61). The model is then completely specified by the polar angles $\lambda_{12}$ & $\lambda_{23}$ of the vectors away from the $\hat{z}$ axis, and by the rotations $\chi_{12}$ & $\chi_{23}$ away from the $zx$-plane.

(ii) *Nearly antiparallel DM vectors*: The results for the last example suggest that we study the case where both DM vector pairs are now *antiparallel*, and then either allow the directions of each pair to vary, or to weakly break the exact antiparallel condition for one or both of them. Thus, in what follows, we will begin by assuming the situation depicted in Fig. 12, in which

$$\hat{\mathbf{d}}_{12} = -\hat{\mathbf{d}}_{34} \quad \& \quad \hat{\mathbf{d}}_{23} = -\hat{\mathbf{d}}_{41}, \qquad (61)$$

but where we allow the relative orientations of the two pairs, associated with the weak and strong links, to vary. We note that X-ray crystallography at 100 K suggests that this configuration is unphysical; however, there are no low-$T$ X-ray results yet, and these may change our understanding of the structure. In any case, the results of the following study are intriguing and informative.

We first study a rather simple case, for which

$$\chi_{\mu\nu} = 0, \qquad (62)$$

so that the DM vectors remain in the $zx$-plane. However, we will allow one of the two DM vector pairs to rotate a small angle away from the other. We do this by fixing

$$\lambda_{23} = 0 \quad (\text{i.e., } \hat{\mathbf{d}}_{23} = -\hat{\mathbf{d}}_{41} = \hat{z}), \qquad (63)$$

but allowing $\hat{\mathbf{d}}_{12} = -\hat{\mathbf{d}}_{34}$ to rotate a small angle away from $\hat{z}$, so that $\lambda_{12}$ is finite. The results are very suggestive (see Fig. 13). One sees again a rapid shift in transverse field of the functional form of the tunneling amplitude, with a concomitant shift of the Berry phase minima. Remarkably, only a 4° tilting of $\lambda_{12}$ away from the vertical is enough to shift the $k = 1(A)$ tunneling amplitude curve by 0.06 T away from its position when $\lambda_{12} = 0$ (and away from the $k = 1(S)$ curve), which is very nearly the shift seen in the experiments.

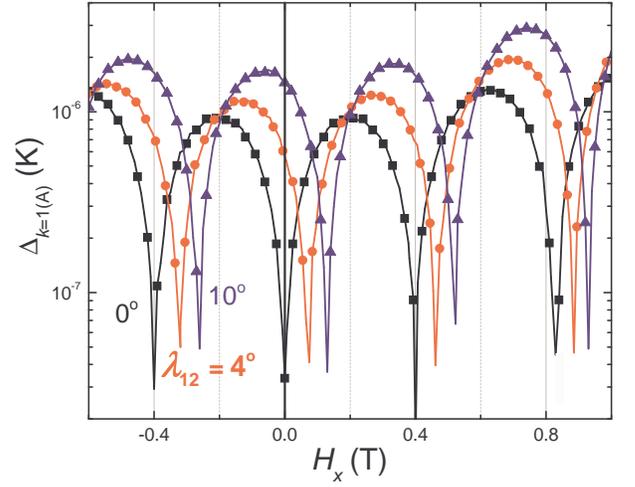

FIG. 13: (Color on-line) Calculated tunneling amplitude for the $k = 1(A)$ resonance, as a function of $H_x$, for the tetramer with broken inversion symmetry (the 'antiparallel case', see Fig. 12). We assume the constraints (61)-(63), so that $\hat{\mathbf{d}}_{23} = -\hat{\mathbf{d}}_{41} = \hat{z}$, but $\hat{\mathbf{d}}_{12}$ is allowed to rotate away from $\hat{z}$. The results are shown for several small rotation angles $\lambda_{12}$ away from $\hat{z}$, within the $zx$-plane (so that $\chi_{12} = 0$).

Now suppose we allow a rotation of the DM vectors out of the $zx$-plane. To do this we fix

$$\lambda_{12} = 4° \quad \& \quad \lambda_{23} = 0, \qquad (64)$$

then allow the vector $\hat{\mathbf{d}}_{12}$ to rotate around the $\hat{z}$-axis, i.e., we vary $\chi_{12}$, as shown in Fig. 14. The result is a gradual rounding of the Berry phase minima, which mimics the effect one observes for conventional Berry phase oscillations as one rotates the external transverse field, $\mathbf{H}_\perp$, away from the hard $\hat{x}$-axis. Note however that this rotation of the DM vectors out of the $zx$-plane also shifts the positions of the $k = 1(A)$ Berry phase minima back towards the $k = 1(S)$ configuration and, once $\chi = 90°$, the reverse shift is complete so that the minima of $k = 1(S)$ and $k = 1(A)$ again coincide.

It is worth stressing that there is no reason to suppose that the tetramer DM vectors must lie in the $zx$-plane. The simple rules given by Moriya [83] allow us to say something about what these directions might be. Recalling that if a mirror symmetry plane bisects the line separating the two spins, then the DM vector is restricted to lie within this plane. One can speculate that since this symmetry exists approximately in the 2 'weak bonds' $\hat{\mathcal{J}}_{12}$ and $\hat{\mathcal{J}}_{34}$, then $\hat{\mathbf{d}}_{12}$ and $\hat{\mathbf{d}}_{34}$ must be approximately restricted to a plane nearly perpendicular to the line joining the two pairs of weakly coupled spins in these bonds. It is interesting that this plane lies 30° away from the hard anisotropy $x$-axis of the molecule (see Fig. 14). Although this hypothesis is rather speculative, it may explain why the data (rounding of $k = 1(A)$ minima) seems to be best fit when $\chi \sim 30°$. This is why this particular angle was chosen to compute the contour plot for the dimer model in Fig. 8.

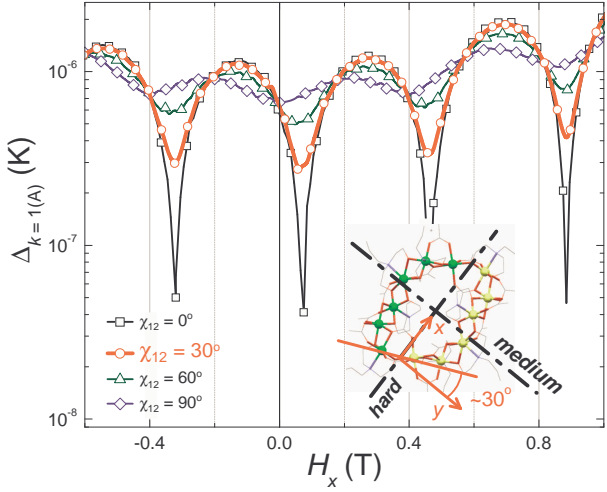

FIG. 14: (Color on-line) Calculated tunneling amplitude for the $k = 1(A)$ resonance, as a function of $H_x$, again for the tetramer with broken inversion symmetry ('antiparallel case'). We again assume that $\hat{\mathbf{d}}_{23} = -\hat{\mathbf{d}}_{41} = \hat{z}$, but now is $\hat{\mathbf{d}}_{12}$ is rotated $\lambda_{12} = 4°$ away from $\hat{z}$ in a plane defined by the angle $\chi_{12}$ (see Fig. 12); the constraint $\hat{\mathbf{d}}_{12} = -\hat{\mathbf{d}}_{34} = \hat{z}$ remains. Results are shown for different values of $\chi_{12}$ (compare to Fig. 13). The inset shows the relative orientation (a 30° angle) between the molecular transverse anisotropy axes and the 'dimer separation line' between the two pairs of weakly coupled spins (see text).

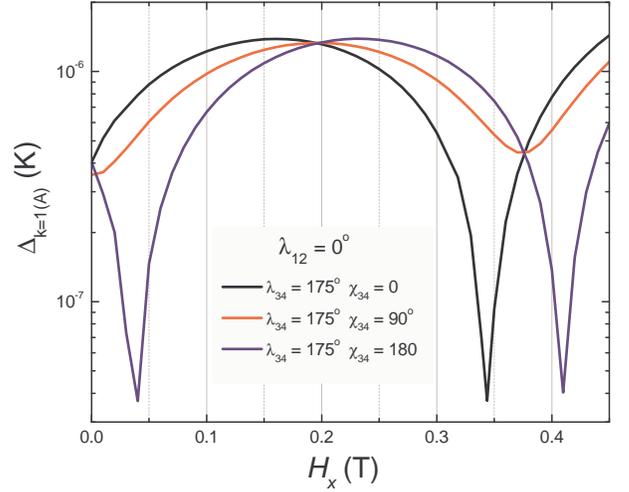

FIG. 15: (Color on-line) The effect of a weak perturbation of the DM vectors, around the strict antiparallel configuration (Fig. 12), on the tunneling amplitude $\Delta_{k=1(A)}$. The vectors are fixed according to eqn. (65), and we show results for three values of $\chi_{34}$.

The most striking conclusion from these results for the tetramer model is that if we wish to find results for the tunneling amplitudes, as a function of transverse field, that look anything like the experiments, then we need to make the drastic assumption that DM vectors on opposite sides of the molecule are oriented nearly antiparallel, representing a completely unphysical distortion from the inversion symmetric case (which requires them to be parallel).

To complete the analysis of this antiparallel case, we need to look at the stability of these results under weak perturbations (disorder) around the strictly antiparallel configuration defined by (61). This was done by fixing

$$\lambda_{12} = 0 \quad \& \quad \lambda_{34} = 175°, \qquad (65)$$

i.e., a small 5° misalignment in one of the two DM vector pairs. We then varied the rotation plane angle $\chi_{34}$. The results are shown in Fig. 15 for three different values of $\chi_{34}$. This behavior is roughly what one might expect: the positions of the Berry phase minima are highly sensitive to $\chi_{34}$, and the rotation also produces a rounding of the minima. More importantly, the minima appear symmetrically on either side of the $k = 1(S)$ positions for angles ranging from 0 to 180°. Consequently, truly random disorder would wipe out the asymmetry in the resulting pattern of $k = 1(A)$ minima about $H_x = 0$.

### V.B.3: Rotation of external field

We have seen previously in the context of the dimer model that the effect of a rotation of the applied field away from the $x$-axis can be compensated by a rotation of the DM vector $\mathcal{D}_{12}$ out of the hard $zx$-plane, allowing a recovery of the Berry phase minima. An obvious question is how this works in the tetramer model. The answer is rather startling. At first glance one might expect to find exactly the same behavior for the strong coupling limit, where $\bar{\mathcal{J}}_S >> \bar{D}_\mu, \bar{K}_\mu, \bar{\mathcal{J}}_w$. However, this is not the case: the results are quite different. To see this, we fix the DM vectors so that

$$\hat{\mathbf{d}}_{12} = -\hat{\mathbf{d}}_{34} \quad \& \quad \hat{\mathbf{d}}_{23} = -\hat{\mathbf{d}}_{41}$$
$$\lambda_{12} = 4° \quad \& \quad \lambda_{23} = 0$$
$$\chi_{12} = \chi_{23} = 30°, \qquad (66)$$

i.e., the antiparallel configuration, but with a 30° rotation away from the $zx$-plane. The resulting tunneling gap is displayed as a contour plot in Fig. 16, as a function of the magnitude and direction of the applied transverse field $\mathbf{H}_\perp$. We see that the behavior is far more rich than found for the simple dimer model in Fig. 8. We make no attempt to give a complete discussion of these results here, which are very complex (the locations of Berry phase minima are extremely sensitive to small

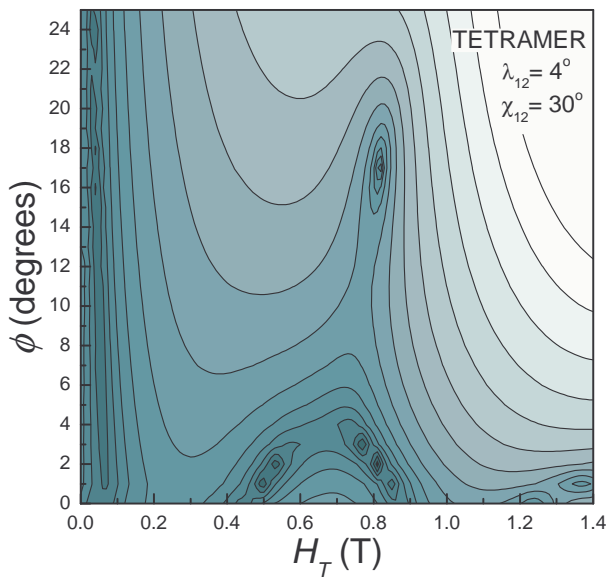

FIG. 16: (Color on-line) Contour plot of the calculated tunnel splitting for the antisymmetric $k = 1(A)$ resonance, in the tetramer model (antiparallel case), with parameters specified by (66). The data are displayed as a function of the strength and direction of the transverse field, $H_\perp$, where the angle $\phi$ is measured relative to the $x$-axis.

variations in the orientations and magnitudes of the coupling parameters).

### V.B.4: Summary of the tetramer model

Even though we have hardly scratched the surface of the tetramer model, the results are sufficiently complex that it is useful to briefly summarize them.

The first main result is that, unsurprisingly the inversion-symmetric tetramer behaves exactly like the inversion-symmetric dimer in the strong coupling limit where $\hat{\mathcal{J}}_S \gg \hat{\mathcal{J}}_w$, i.e., the antisymmetric tunneling transitions are strictly forbidden. Relatively weak perturbations of the inversion symmetry, such as small misalignments ($\sim 1°$) of the DM vectors, 'switch on' the antisymmetric resonances (e.g. $k = 1(A)$). However, the pattern of Berry phase minima, as a function of $H_\perp$, remains unshifted from that of the $k = 0$ Berry phase oscillations for these weak perturbations. Generation of the experimentally observed 'asymmetric $H_\perp$ shift' for $k = 1(A)$ requires a very substantial perturbation: the weak DM pair must have an antiparallel arrangement (as opposed to the parallel arrangement required by the resonance inversion symmetry). Weak perturbations about this antiparallel case (including rotations away from the $zx$-plane) do yield results that are similar to experiment. We emphasize again that if the X-ray data taken at 100 K are still valid at low $T$, then such a configuration is most likely unphysical for the $Mn_{12}$ ring. We discuss this point more in the concluding section.

## VI: CONCLUDING REMARKS

If one takes the preceding simulations seriously, then, as noted already several times, the existence of a net DM vector in a system which is manifestly inversion-symmetric poses a real problem. In their later work, which essentially repeated ours, Wernsdorfer et al. [71, 75] presented the same kinds of simulation, with a comparable DM vector. However, they argued that this was compatible with the symmetry of the $Mn_{12}$ wheel molecule, citing papers [97] in which, they claimed, inversion symmetric systems have finite net DM vectors. This argument is incorrect. As we noted in sections II and III above, and as is in any case well known, the existence of a DM vector for a given exchange or superexchange bond depends only on whether there is inversion symmetry about the center *of that bond*. Thus the DM vectors $\mathbf{D}_{ij}$ in each $Mn \cdots Mn$ bond of the $Mn_{12}$ wheel molecule are in general finite, but the DM vector $\mathcal{D}_{12}$ between the two dimer halves, in the dimer *ansatz*, *has to* be zero if the molecule itself is inversion symmetric. This means of course that a proper sum over the internal DM vectors $\mathbf{D}_{ij}$ of the molecule, taking into account the direction and sense of each bond, does indeed have to give a 'net DM vector' of zero. Indeed, as noted in section II.A.2, in the inversion-symmetric wheel molecule there are six pairs of $\mathbf{D}_{ij}$, with each pair related by inversion symmetry, and hence equal; this immediately gives $\mathcal{D}_{12} = 0$ (NB: none of the results in ref. [97] are incompatible with these remarks).

It is quite apparent that we need some other way to explain the existence of the antisymmetric tunneling transitions. For this reason, we have given considerable attention to the possible consequences of disorder in our analysis of the tetramer model. In particular, we showed that the antisymmetric resonance is 'switched on' by perturbations which cause a local breaking of the inversion symmetry. This also represents one of the many explanations given by Wernsdorfer et al. [71, 75] for the appearance of the antisymmetric transition. The problem is that disorder typically generates a random distribution of perturbations. In the dimer picture this would involve a random symmetric distribution over different molecules for the single DM vector $\mathcal{D}_{12}$; and in the tetramer model, a random distribution of misalignments (tilts) of the four DM vectors away from the directions required by the inversion symmetry, as well as some randomness in the magnitude of these vectors. As noted when discussing both the dimer and tetramer models, rotations of the DM vectors have a profound influence on the transverse field behavior of the Berry phase minima associated with the antisymmetric resonance. Consequently, any randomness associated with any disorder would lead to a complete elimination of the quantum oscillations observed in the experiment. This fact, alone, seems to rule out any ex-planation in terms of random disorder.

One might try arguing that the disorder is not random, but discrete, as was found for $Mn_{12}$-acetate. However, several factors seem to rule out this possibility as well. In particular, we have shown that a very substantial perturbation of the inversion symmetry, requiring nearly antiparallel DM vectors on the weak bonds, is needed in order to account for the experimentally observed behavior. If one assumes that the structure of the $Mn_{12}$ ring molecule found in X-ray studies at 100 K is also valid at low $T$, this seems extremely unlikely. The antiparallel arrangement of DM vectors would likely correspond to a very significant distortion of the $Mn_{12}$ ring molecule. One way to see this is by considering how the DM interactions $D_{ij}$ arise on the individual bonds in the $Mn_{12}$ ring. Consider in particular the weak "bent" $Mn\cdots O\cdots Mn$ pathways between the 2 dimer halves (see Fig. 1). Because of the inversion symmetry within the full wheel, the bonds on opposite sides bend in the same way, i.e., the O atom is displaced outwards from the wheel center on both sides. It is this symmetry that ensures that the DM vectors associated with those two bonds are *parallel*. In order for the DM vectors to be *antiparallel*, the O atom in one of the two bonds would be expected to buckle *inwards* by roughly the same amount that the other buckles *outwards*. Clearly, this would represent a very significant distortion, which should show up very clearly in the probability ellipsoids deduced from analysis of the X-ray scattering data. This is not what is seen in the 100 K measurements, i.e. the positions of the O atoms are very clearly defined.

However, as already noted above, one should also consider the possibility that there may be a phase transition below $\sim 100$ K, in which a change of symmetry occurs, so that *each molecule* loses inversion symmetry. Note that this is not a disorder effect; the transition would result in a new crystal in which each molecule has the same, non-inversion symmetric shape. An example that might explain the experiments would involve a buckling of the $Mn\cdots O\cdots Mn$ bonds associated with the weakest exchange links, as described above. At the moment this is no more than a hypothesis, which needs to be tested. However, we note that at present there is no evidence against it. No X-ray measurements have been performed at low temperatures. Furthermore, although low-T specific heat and susceptibility measurements have been reported [80], they were performed on powders, and there is a large scatter in the data which may obscure any possible phase transition.

Finally, it is worth taking a step back and asking whether inclusion of DM terms introduces any fundamentally new physics. If a given molecule is centrosymmetric, then parity is conserved (ignoring for the moment issues related to disorder, dynamics, etc.). In such a situation, spin states of opposite parity cannot mix. However, if the inversion symmetry is broken, parity is no longer conserved. Thus, one need not invoke DM interactions to generate the required mixing: ordinary single-ion anisotropy (i.e., $D$ and $E$) will do the job, provided one or more of the exchange links within the molecule is relatively weak [98]. The typical single-ion anisotropies associated with Jahn-Teller distorted $Mn^{3+}$ ions are in fact considerably stronger than the presumed weakest exchange links ($\mathcal{J}_w$) in the $Mn_{12}$ wheel molecule. Consequently, one may expect considerable mixing, e.g., between spin $S \sim 6$ and $S \sim 7$ states, due to the purely symmetric interactions [69, 98, 99]. In other words, provided the inversion symmetry is somehow broken, then perhaps DM interactions are not needed at all.

In principle it would of course be very interesting if one were able to construct a more exact model in which the spin-orbit anisotropy (both the local and exchange contributions) is treated rigorously on all of the ions and bonds in the molecule; one could then analyze the influence of each interaction on the molecular tunneling, upon breaking the inversion symmetry. Such an analysis is well beyond contemporary computational methods: the $Mn_{12}$ wheel molecule without any imposed symmetry would require $> 150$ adjustable parameters (not to mention the enormous dimension of the Hamiltonian matrix). It is notable, however, that for low-nuclearity (simpler) Mn systems for which detailed characterizations have been performed to date, exceptionally good agreement is usually found without the need to invoke DM interactions [69, 100]. This is probably because the DM terms are far weaker than the single-ion anisotropies ($\mathcal{D}_{ij} \sim (\mathcal{J}_{ij}/\lambda_{so}) \times D$, where $\lambda_{so}$ is the spin-orbit coupling energy). Therefore, the effects of the DM interaction are unmeasurably small, even though we note that the $\mathcal{D}_{ij}$ mix spin states in a lower order of perturbation compared to the single-ion terms ($D$, $E$, etc.) [101]. It is only in spin-1/2 systems (e.g. $Cu^{2+}$) that the DM interaction has clearly recognizable consequences, where it represents the only possible source of anisotropy. Stated differently, for $S = 1/2$, it is the only allowed interaction that is capable of mixing spin multiplets, thereby yielding rich new physics beyond that predicted on the basis of a pure Heisenberg description [97, 101].

## ACKNOWLEDGEMENTS

We acknowledge discussions with Eduardo Mucciolo and Nick Bonesteel. The research of E. del Barco was supported by NSF (DMR-0747587). The research of S. Hill was supported by NSF (DMR-0804408 and CHE-0924374). The research of P.C.E. Stamp was supported by NSERC and CIFAR in Canada. The research of I.S. Tupitsyn was supported by PITP.